# A comprehensive study on causal discovery between degradation paths

Shi-Shun Chen [a, b], Shuai Gao [a], Xiao-Yang Li [a,*], Enrico Zio [b, c]

[a] School of Reliability and Systems Engineering, Beihang University, Beijing, China

[b] Energy Department, Politecnico di Milano, Milan, Italy

[c] Centre de Recherche sur les Risques et les Crises (CRC), MINES Paris-PSL University, Sophia Antipolis, France

Email:
css1107@buaa.edu.cn (Shi-Shun Chen)
gs0828@buaa.edu.cn (Shuai Gao)
leexy@buaa.edu.cn (Xiao-Yang Li)
enrico.zio@polimi.it, enrico.zio@mines-paristech.fr (Enrico Zio)

Corresponding author*: Xiao-Yang Li

## Highlights

- Pairwise causal relationships between degradation paths are investigated.
- A causal discovery strategy based on degradation increments is proposed.
- Five different types of non-temporal causal discovery techniques are compared.
- The impact of degradation process characteristics on causal discovery is evaluated.
- Stable-PC and GES are recommended for causal discovery between degradation paths.

## Abstract

Existing studies indicate that complex system degradation is characterized by degradation of multiple dependent parameters. Capturing the dependencies is crucial for accurate degradation modeling and effective degradation control. This work aims to uncover these dependencies through causal analysis, focusing on pairwise causal discovery. Firstly, considering the steady-state characteristic of physical dependencies between parameters, a causal discovery strategy using degradation increments is proposed combined with non-temporal causal discovery techniques. Then, five types of non-temporal causal discovery techniques, including constraint-based, score-based, functional causal model-based, gradient-based and the emerging ordering-based technique, are selected as benchmark methods to identify the most suitable approach. Numerical studies based on Wiener process are first conducted to investigate the method effectiveness on both independent and causally dependent degradation paths. Additionally, sensitivity analysis is performed to evaluate how degradation process characteristics affect the accuracy of causal discovery. Then, two engineering applications are given to show the practical applicability of the approach, including a second-order multiple-feedback band pass filter and a turbofan engine. Our findings indicate that the proposed strategy, which uses degradation increments, outperforms methods that rely on raw degradation data. Among all evaluated techniques, stable Peter-Clark and greedy equivalence search exhibit robust and accurate performance across both numerical and engineering cases, which are recommended for causal discovery between degradation paths. The code is available on GitHub: https://github.com/dirge1/causal_deg_data.

# 1 Introduction

Analyzing degradation data plays an essential role in reliability analysis and remaining useful life (RUL) prediction. A majority of research has focused on the degradation of a single parameter considered to represent overall system performance [1-3]. However, in practical applications, the performance of engineered systems is always characterized by multiple parameters [4-6]. Since the performance parameters are always dependent due to inherent physical principles, their degradation processes are also dependent [7]. To achieve precise degradation modeling and effective degradation control, it is essential to capture the dependencies between degradation processes.

To describe the dependencies between degradation processes, correlation analysis is primarily employed, which can be generally classified into three categories. The first focuses on capturing the correlations between parameters in the degradation models of different variables, such as correlations of drift coefficients [8], random effects [9], memory effects [10] and noise [11]. The second category employs copula functions to quantify the dependence in the probability distributions of different degradation paths, such as the correlation of first-passage time distributions (i.e., lifetime distributions) [12, 13] and the correlation of degradation increment distributions [14, 15]. The third category posits that the degradation rate of one variable is influenced by the degradation states of other variables, with their relationships described using a state-space correlation coefficient matrix [16-18]. All these approaches have demonstrated effectiveness in degradation modeling and RUL prediction of multivariate degradation systems.

However, correlation does not imply causation [19]. For example, the exhaust gas temperature (EGT) [20] and the fuel flow (FF) [21] are two critical parameters describing engine performance. During engine operation, an increase in FF intensifies combustion, raises gas temperature, and ultimately increases EGT. Therefore, FF has a causal impact on EGT and these two parameters are highly correlated. Since correlation-based methods do not capture causal directionality, they may incorrectly suggest that a decrease in EGT always indicates an increased likelihood of FF decline. Nevertheless, FF determines the engine thrust and is determined by operating requirements. A decrease in EGT will not lead to FF decline, which correlation-based methods fail to recognize. Hence, identifying the causal relationships among degradation paths is crucial for accurately comprehending their dependencies.

Causal discovery is a technique for uncovering causal relationships among variables using observation data under specific assumptions, and it has already been successfully applied in fault detection [22-24] and fault diagnosis tasks [25-27]. By employing causal discovery techniques, it becomes possible to identify the inherent causal relationships among degradation paths, thus addressing the limitations of correlation-based approaches.

However, application of causal discovery techniques to degradation paths still faces significant challenges. Firstly, degradation data struggle to meet the assumptions required by causal discovery techniques. Causal discovery techniques are generally divided into temporal methods (i.e., methods for time-series data) and non-temporal methods (i.e., methods for independent and identically distributed (IID) data) [28]. In existing studies, temporal causal discovery methods are usually applied to industrial time series data for fault detection and fault diagnosis, including Granger causality (GC) [27, 29], transfer entropy (TE) [30, 31] and convergent cross mapping (CCM) [32, 33]. Temporal causal discovery methods assume that time series data involve the dynamics of variable relationships. In their framework, a causal link from variable $X_1$ to variable $X_2$ implies that the current state of $X_2$ is affected by the historical states of $X_1$. However, degradation data are typically recorded in a steady state. For dependent degradation processes, the degradation state of a variable is always influenced by the current states of other variables rather than their historical states, like the causal impact of FF on EGT. As a result, despite being time series data, degradation data fail to fit the assumptions of temporal causal discovery methods, as shown in Fig. 1, and therefore the non-temporal causal discovery methods can be more suitable. Although some studies have applied non-temporal methods to steady-state observation data for fault detection [22-24], they do not account for performance degradation. In their research, variables are typically IID and satisfy the assumption of non-temporal causal discovery methods. However, since degradation data reflect performance changes over time, they violate the IID assumption, posing challenges for causal discovery based on the raw degradation data.

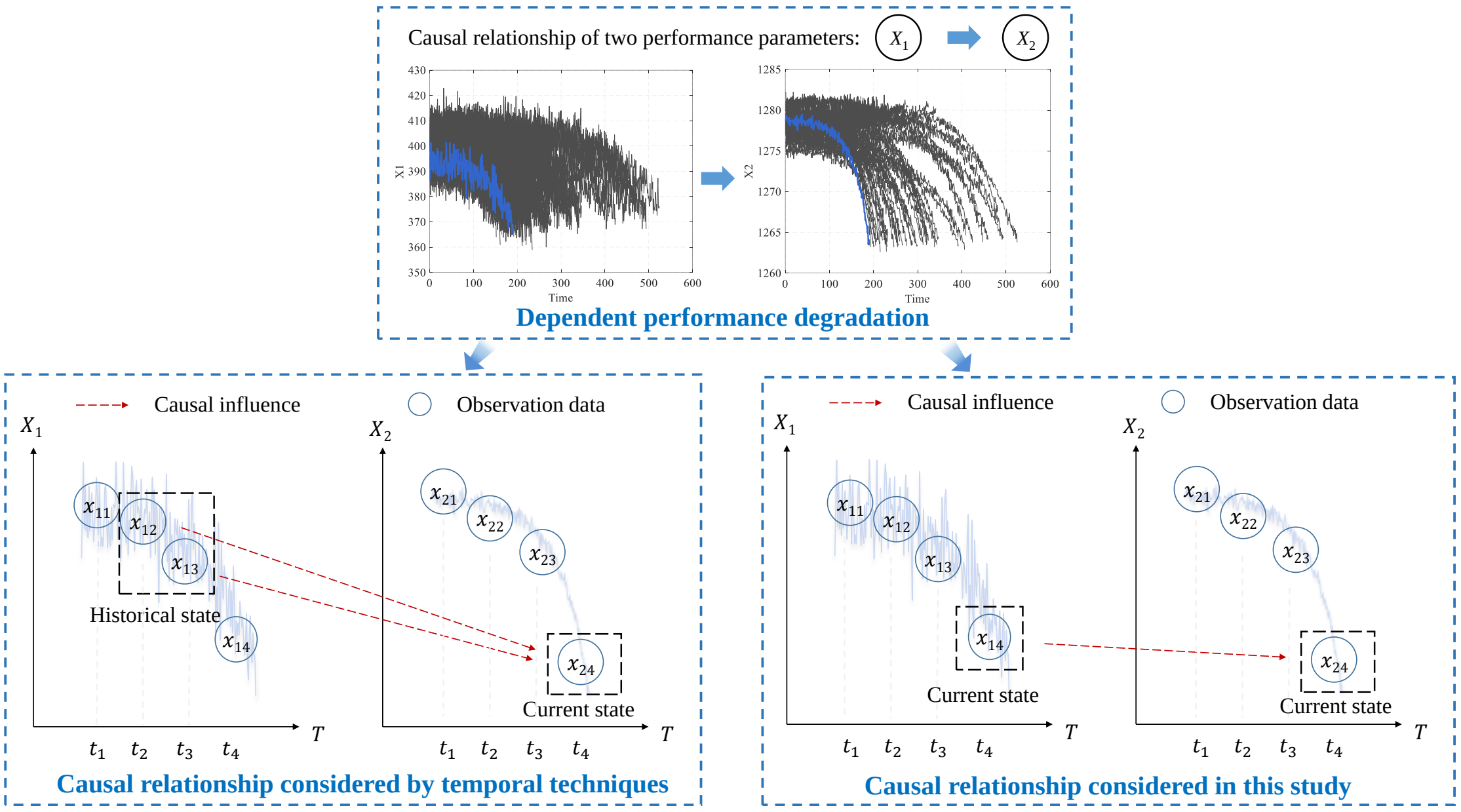

Fig. 1 Schematic of the causal relationship in degradation data from $X_1$ to $X_2$.

Secondly, the correctness of causal graphs and the applicability of causal discovery techniques have been overlooked in current research. With the advancement of causal discovery techniques, a vast number of causal discovery techniques have emerged. However, existing studies employ causal discovery techniques without assessing their suitability. Instead, they have primarily focused on developing graph-based deep learning algorithms for fault detection and diagnosis using causal graphs, especially the application of graph neural networks [22-27]. Although the superiority of using causal graphs has been validated through comparisons with correlation graphs and ablation experiments, the correctness of the causal graphs themselves has yet to be confirmed.

In response to the above challenges, this study aims to explore effective strategies and techniques for uncovering causal relationships between degradation paths. First, a causal discovery strategy is presented using degradation increments combined with non-temporal causal discovery techniques. Degradation increments offer a way to satisfy the IID assumption while preserving the causal relationships between variables. Then, five types of non-temporal causal discovery techniques are introduced, including constraint-based, score-based, functional causal model-based, gradient-based and the emerging ordering-based approach [28]. These comparisons aim to identify the most suitable technique for discovering causal relationships between degradation paths. For case studies, a numerical case is first presented to verify the effectiveness of the proposed method for both independent and causally dependent degradation paths, and sensitivity analysis is carried out to analyze the impact of various factors on the causal discovery results. The Wiener process is utilized to simulate degradation trajectories due to its capability of describing most of degradation data [34]. Next, two engineering applications are given to show its practical applicability, including a second-order multiple-feedback band pass filter and a turbofan engine.

In this study, we focus on pairwise causal discovery, as pairwise causality forms the fundamental building block for complex causal networks. The exploration of causal discovery among multiple degradation paths is reserved for future work due to its significantly greater complexity [35]. To the best of our knowledge, this is the first study specifically focused on causal discovery between degradation paths, which can be the foundation of multivariate degradation modeling, reliability analysis and degradation control considering inherent causal relationships.

The main contributions of this paper can be summarized as follows:

- A causal discovery strategy using degradation increments combined with non-temporal causal discovery techniques is presented for degradation paths, mitigating the negative impact of trends in raw degradation data while preserving causal relationships between variables.
- The effectiveness of five types of non-temporal causal discovery techniques are compared in both numerical and engineering cases, so as to identify the most suitable techniques for causal discovery between degradation paths.
- The impacts of causality nonlinearity, degradation nonlinearity, random effects, measurement error and diffusion coefficient on causal discovery accuracy are examined through sensitivity analysis, which can provide guidance for assessing the feasibility of causal discovery between degradation paths.

The overview of this study is illustrated in Fig. 2.

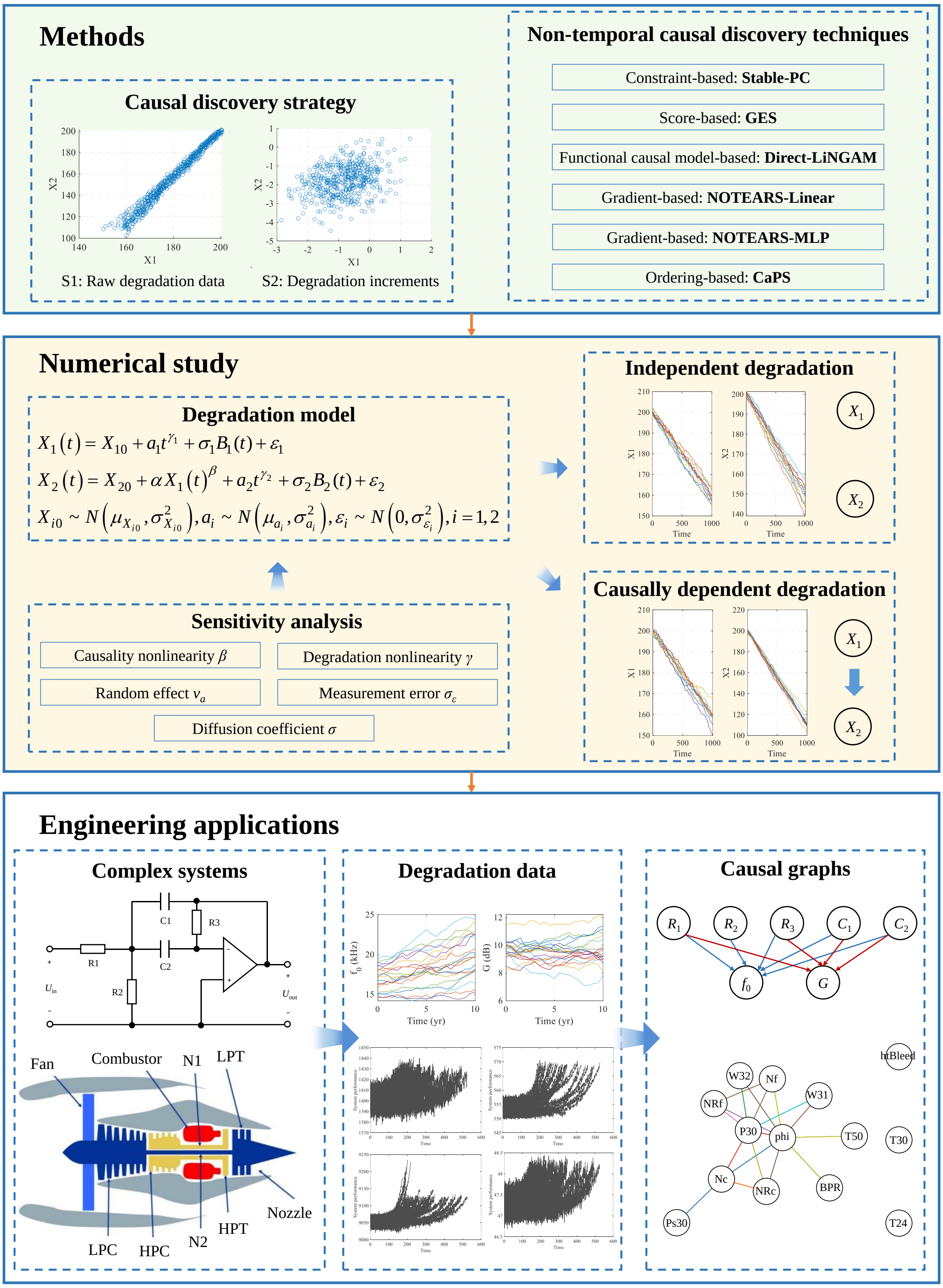

Fig. 2 Overview of the study.

# 2 Methods

## 2.1 Non-temporal causal discovery techniques

Causal discovery techniques are generally divided into temporal methods and non-temporal methods [28]. As mentioned earlier, the dependent degradation processes described in Eq. (5) fail to satisfy the assumptions of temporal causal discovery methods. Therefore, non-temporal methods should be employed. This subsection introduces the non-temporal causal discovery techniques used in this study.

Notably, the limitations of temporal methods for degradation paths are also examined in numerical case studies.

Non-temporal causal discovery techniques are commonly categorized into four types [28]: constraint-based, score-based, functional causal model-based and gradient-based. Additionally, there exists several methods employing specialized techniques to search for causal graphs. In this study, four abovementioned categories of causal discovery techniques are covered, as well as a newly developed ordering-based approach.

First, we give the definitions of several essential terms related to causal discovery.

*Definition 1* (directed acyclic graph): A directed acyclic graph (DAG) is a finite directed graph that contains no directed cycles. In causal discovery, DAGs are used to represent causal relationships among variables, where nodes represent variables, and directed edges from one node to another indicate causal influence. The acyclic property ensures that no feedback loops exist in the causal structure.

*Definition 2* (skeleton): A skeleton of a DAG is its undirected version, obtained by replacing all directed edges with undirected edges while preserving connectivity. The skeleton captures variable dependencies but does not specify the direction of causal relationships.

*Definition 3* (V-structure): A V-structure is a specific pattern in a DAG where two parent nodes point to a common child node without being directly connected, i.e., $A \rightarrow C \leftarrow B$, where $A$ and $B$ are not adjacent.

*Definition 4* (Markov equivalence class): A Markov equivalence class (MEC) is a set of DAGs that encode the same conditional independence relationships. Two DAGs belong to the same MEC if and only if they share the same skeleton and the same V-structure.

Then, the causal discovery techniques adopted in this study are described as follows:

**Stable Peter Clark (Stable-PC)** [36]: Stable-PC is a **constraint-based** causal discovery technique that infers causal structures using conditional independence (CI) tests. It is an improved version of the PC algorithm, which constructs a DAG by initially forming a fully connected undirected graph, and then progressively removing edges through CI tests. The main limitation of the original PC algorithm is its sensitivity to the order of CI tests, which can lead to unstable results. Stable-PC addresses this issue by ensuring that edges are removed in a consistent and order-independent manner, providing more robust and reliable causal structures.

**Greedy equivalence search (GES)** [37]: GES is a widely used **score-based** causal discovery method that learns causal structures by optimizing a predefined score function over MECs. The algorithm consists of two phases: a forward phase, where edges are greedily added to maximize the score, and a backward phase, where redundant edges are removed to refine the structure. The score function typically balances data likelihood and model complexity, with commonly used choices like the Bayesian information criterion (BIC) [38].

**Direct linear non-Gaussian acyclic model (Direct-LiNGAM)** [39]: Direct-LiNGAM is a **functional causal model-based** causal discovery method that identifies causal relationships by assuming a linear non-Gaussian acyclic model. Unlike CI test-based methods, Direct-LiNGAM exploits the independence of structural noise terms to infer causal directionality. It directly estimates the causal ordering of variables using independent component analysis (ICA), and then removes spurious edges via a regression-based approach. By leveraging the non-Gaussian assumption, Direct-LiNGAM allows for identifying unique causal structures.

**Non-combinatorial optimization via trace exponential and augmented lagrangian for structure learning (NOTEARS)** [40]: NOTEARS is a **gradient-based** causal discovery method that reformulates DAG learning as a continuous optimization problem. It introduces a differentiable acyclicity constraint, allowing for efficient structure learning using standard gradient-based optimization techniques. **NOTEARS-Linear** [40] is the original version of NOTEARS, where the causal relationships between variables are assumed linear. It models each variable as a linear function of its parents and optimizes the adjacency matrix using least squares loss while ensuring that the DAG constraints are met. **NOTEARS-MLP** [41] extends NOTEARS-Linear to nonlinear causal relationships by replacing the linear functional form with a multi-layer perceptron (MLP), which is able to capture complex nonlinear dependencies while maintaining the computational efficiency of gradient-based optimization.

**Causal Discovery with Parent Score (CaPS)** [42]: CaPS is an **ordering-based** causal discovery method that learns causal structures by first determining a topological ordering of variables, and then constructing a DAG based on a novel parent selection strategy. Unlike traditional ordering-based approaches that assume either purely linear or nonlinear causal relationships, CaPS introduces a unified criterion to identify leaf nodes using the Hessian expectation of the data log-likelihood, allowing it to handle both linear and nonlinear causal relations effectively. After determining the causal ordering, CaPS refines the structure by employing a parent score metric, which quantifies the average causal effect strength to improve edge selection and pruning. This metric helps to correct misclassified edges and enhance the final graph structure. By integrating pre-pruning and edge supplementation techniques, CaPS improves computational efficiency while maintaining accuracy.

The methods Stable-PC, GES, Direct-LiNGAM, NOTEARS-Linear and NOTEARS-MLP are widely recognized as mature approaches in causal discovery, demonstrating effectiveness in applications such as health management and risk analysis [22-24, 43-45]. Additionally, CaPS is a novel causal discovery technique that has shown promising potential in uncovering complex causal relationships. Therefore, these six methods are selected for benchmark study in this research.

For the implementation of these methods, we use gcastle [46], a causal discovery toolbox developed by Huawei, to implement the first five techniques. For the ordering-based approach, we employ the source code provided by the authors [42]. The hyper-parameters of each method, along with their descriptions and values, are shown in Table 1.

Table 1 Descriptions and values of the hyper-parameters for each causal discovery technique.

| Methods | Descriptions | Values |
|---|---|---|
| Stable-PC | Significance level of CI tests | 0.05 |
| | Method for CI tests | Fisher's Z test |
| GES | Method for score function | BIC |
| Direct-LiNGAM | Method for CI tests | Pairwise likelihood ratio |
| | Threshold for causality determination | 0.1 |
| NOTEARS-Linear | L1 regularization strength | 0.1 |
| | Maximum number of iterations | 100 |
| | Tolerance for DAG constraint | 1e-8 |
| | Maximum penalty parameter | 1e16 |
| | Threshold for causality determination | 0.1 |
| NOTEARS-MLP | L1 regularization strength | 0.01 |
| | L2 regularization strength | 0.01 |
| | Maximum number of iterations | 100 |
| | Tolerance for DAG constraint | 1e-8 |
| | Maximum penalty parameter | 1e16 |
| | Number of hidden layers | 10 |
| | Threshold for causality determination | 0.1 |
| CaPS | Threshold for pre-pruning | 50 |
| | Threshold for edge supplementation | 50 |
| | Threshold for causal additive model pruning | 0.1 |

## 2.2 Causal discovery strategy for degradation paths

Before introducing causal discovery strategies, we first define the notation of degradation data. Let $x_{lij}$ represent the $j^{\text{th}}$ degradation value of the $l^{\text{th}}$ performance parameter during the test of the $i^{\text{th}}$ sample, where $l = 1, 2, ..., k$, $i = 1, 2, ..., n$, $j = 1, 2, ..., m_{li}$, where $k$ is the number of performance parameters, $n$ is the number of test items, and $m_{li}$ denotes the number of measurements for unit $i$ in response to the $l^{\text{th}}$ performance parameter. The corresponding measurement time is denoted as $t_{lij}$. Without loss of generality, we assume that the measurement times are identical across different samples and performance parameters, i.e., $t_{lij} = t_j$, where $t_j$ represents the measurement time of the $j^{\text{th}}$ degradation value.

For the causal relationships between degradation paths, the following assumptions are considered:

- The causal relationships between degradation paths are exhibited in the steady state, as illustrated in Fig. 1.
- These causal relationships are stable and time-invariant, consistent with assumptions commonly adopted in existing causal discovery studies.

Then, an intuitive strategy for causal discovery is to directly analyze the raw degradation data, since the parameter values at the same observation time for the same sample contain causal information among the variables. This is also the most commonly used strategy in existing studies [22-24]. We denote this strategy as $S_1$, and the data for causal discovery can be expressed as:

$$\mathbf{X}_{S1} = \begin{bmatrix} x_{111} & x_{112} & \cdots & x_{11m_1} & x_{121} & x_{122} & \cdots & x_{1nm_n} \\ x_{211} & x_{212} & \cdots & x_{21m_1} & x_{221} & x_{222} & \cdots & x_{2nm_n} \\ \vdots & \vdots & \ddots & \vdots & \vdots & \vdots & \ddots & \vdots \\ x_{k11} & x_{k12} & \cdots & x_{k1m_1} & x_{k21} & x_{k22} & \cdots & x_{knm_n} \end{bmatrix}^{\mathrm{T}} \tag{1}$$

However, as the system performance degrades over time, its performance parameters exhibit a trend. At this point, the data characteristic in Eq. (1) significantly violates the IID assumption required by non-temporal causal discovery methods, making it changeling to accurately identify causal relationships among degradation paths. To address this, this paper presents a causal discovery strategy based on degradation increments. Let $\Delta x_{lij}$ represents the degradation increment between the $j^{\text{th}}$ and $(j\text{-}1)^{\text{th}}$ measurement of the $l^{\text{th}}$ performance parameter during the test of the $i^{\text{th}}$ sample, which can be calculated as:

$$\Delta x_{lij} = x_{lij} - x_{li(j-1)}, j = 2,3,\ldots,m_{li} \tag{2}$$

Then, we denote the causal discovery strategy based on degradation increments as $S_2$, and the data for causal discovery can be expressed as:

$$\mathbf{X}_{S2} = \begin{bmatrix} \Delta x_{112} & \Delta x_{113} & \cdots & \Delta x_{11m_1} & \Delta x_{122} & \cdots & \Delta x_{1nm_n} \\ \Delta x_{212} & \Delta x_{213} & \cdots & \Delta x_{21m_1} & \Delta x_{222} & \cdots & \Delta x_{2nm_n} \\ \vdots & \vdots & \ddots & \vdots & \vdots & \ddots & \vdots \\ \Delta x_{k12} & \Delta x_{k13} & \cdots & \Delta x_{k1m_1} & \Delta x_{k22} & \cdots & \Delta x_{knm_n} \end{bmatrix}^{\mathrm{T}} \tag{3}$$

From the perspective of degradation increments, the trend in parameter values caused by performance degradation can be largely mitigated, while the degradation increments still retain causal relationship information among variables.

# 3 Numerical study

In this paper, we focus on causal discovery between two degradation paths, for which there are two key scenarios: 1) When two degradation paths are independent, the method should correctly recognize their independence and avoid detecting false causal relationships; 2) When two degradation paths are causally dependent, the method should accurately identify the existence of the causal relationship and determine its direction. Accordingly, the numerical cases are categorized into two parts: 1) Causal discovery for independent degradation paths; 2) Causal discovery for causally dependent degradation paths. These two parts are presented in Sections 3.3 and 3.4, respectively.

## 3.1 Simulated degradation paths based on Wiener process

Since the true causal graphs of real-world degradation data are often unknown, numerical studies are first introduced. This allows analyzing various factors influencing causal discovery through sensitivity analysis, and providing valuable insights for practical applications. As a classic model used for degradation modeling, the Wiener process-based degradation model can effectively represent the degradation paths of most degradation data [34]. Therefore, it is employed as the benchmark model for numerical case studies in this paper.

Initially, we focus on the degradation of a single parameter $X$. Based on the results in the literature, its performance degradation can be generally described as [47-49]:

$$\begin{aligned} & X(t) = X_0 + a\Lambda(t) + \sigma B(t) + \varepsilon \\ & X_0 \sim N\left(\mu_{X_0}, \sigma_{X_0}^2\right), a \sim N\left(\mu_a, \sigma_a^2\right), B(t) \sim N\left(0, t\right), \varepsilon \sim N\left(0, \sigma_\varepsilon^2\right) \end{aligned} \tag{4}$$

where $t$ is the degradation time; $X_0$ is a normally distributed random variable that describes the initial value of $X$ with uncertainties [47]; $a$ is a normally distributed random variable that describes the degradation rate with random effects [48]; $\Lambda(t)$ is the time-scale function that describes the degradation trend; $\sigma$ is the diffusion coefficient; $B(t)$ represents a standard Wiener process (i.e., Brownian motion); $\varepsilon$ is a normally distributed random variable quantifying measurement errors [49]; and $\mu_i$ and $\sigma_i$ represent the mean value and standard deviation of the random variable $i$, respectively.

As mentioned earlier, degradation data are typically recorded in steady state. For multivariate degradation systems, the degradation state of a variable is influenced by the current states of other variables. Then, for a bivariate degradation system with a causal link from $X_1$ to $X_2$, the degradation model can be constructed as:

$$
\begin{aligned}
&X_1(t) = X_{10} + a_1\Lambda_1(t) + \sigma_1 B_1(t) + \varepsilon_1 \\
&X_2(t) = X_{20} + g_2\left(X_1(t)\right) + a_2\Lambda_2(t) + \sigma_2 B_2(t) + \varepsilon_2 \\
&X_{i0} \sim N\left(\mu_{X_{i0}}, \sigma_{X_{i0}}^2\right), a_i \sim N\left(\mu_{a_i}, \sigma_{a_i}^2\right), B_i(t) \sim N(0,t), \varepsilon_i \sim N\left(0, \sigma_{\varepsilon_i}^2\right), i = 1,2
\end{aligned}
\tag{5}
$$

where the definitions of $X_{i0}$, $a_i$, $\Lambda_i$, $\sigma_i$, $B_i(t)$ and $\varepsilon_i$ are similar to those in Eq. (4), with the subscript $i$ indicating that the parameter belongs to the degradation model of $X_i$, and $g_i(\cdot)$ represents the function describing the causal influence of the parent variables of $X_i$ on itself. Notably, apart from the causal influence of $X_1$ on $X_2$, $X_2$ itself also undergoes performance degradation. This corresponds to actual conditions, since different performance parameters may have different degradation modes. Note that the bivariate degradation model in Eq. (5) can be easily extended to consider more variables with inherent causal relationships.

### 3.2 Evaluation metrics

To evaluate the performance of the causal discovery techniques considered and ensure robustness, we define the exact match rate (EMR) as:

$$
\text{EMR} = \frac{1}{N}\sum_{i=1}^{N} I(R_i) \tag{6}
$$

where

$$
I(R_i) = \begin{cases} 1, & \text{if } R_i = G \\ 0, & \text{otherwise} \end{cases} \tag{7}
$$

is an indicator function, $N$ is the total simulation times; $R_i$ represents the causal graph obtained from the causal discovery for the $i^{\text{th}}$ set of degradation paths; and $G$ denotes the true causal graph.

### 3.3 Causal discovery between independent degradation paths

#### 3.3.1 Numerical experiment design

For independent degradation paths, the parameter settings of Eq. (5) are shown in Table 2. Specifically, the time scale functions are given by $\Lambda_1(t) = t^{\gamma_1}$ and $\Lambda_2(t) = t^{\gamma_2}$, which have exhibited broad applicability in capturing the degradation trajectories of various kinds of products [50-52]; measurement time interval $\Delta t$ for both degradation paths across all samples are set as 20, and the total measurement time for both paths is set as 1000, with $m = 51$ for each path; $g_2\left(X_1(t)\right) = 0$ since there is no causal relationship; and $v_{a_i}$ represents the variable coefficient of the random effect in the degradation model of $X_i$, calculated by:

$$
v_{a_i} = \sigma_{a_i} / \mu_{a_i} \tag{8}
$$

Table 2 Parameter settings of independent degradation paths

| Parameters | Values | Parameters | Values |
|---|---|---|---|
| $\mu_{X_{10}}$ | 200 | $\mu_{X_{20}}$ | 200 |
| $\sigma_{X_{10}}$ | 1 | $\sigma_{X_{20}}$ | 1 |
| $\mu_{a_1}$ | -0.04 | $\mu_{a_2}$ | -0.05 |
| $v_{a_1}$ | 0.05 | $v_{a_2}$ | 0.05 |
| $\sigma_1$ | 0.1 | $\sigma_2$ | 0.1 |
| $\sigma_{\varepsilon_1}$ | 0.4 | $\sigma_{\varepsilon_2}$ | 0.4 |
| $\gamma_1$ | 1 | $\gamma_2$ | 1 |
| $\Delta t$ | 20 | $m$ | 51 |

In practical engineering applications, the number of degradation samples needed for causal discovery is a critical factor. Moreover, it is essential to evaluate the convergence of the causal discovery methods with increasing sample size. Hence, we investigate the effect of sample size on the causal discovery results across all analysis. Considering that small sample situations are common in engineering applications, we set the sample size range from one to ten. Twenty sets of degradation paths are simulated independently ($N$ = 20) and causal discovery is performed for each to evaluate robustness. An example of independent degradation paths with ten sample size is shown in Fig. 3.

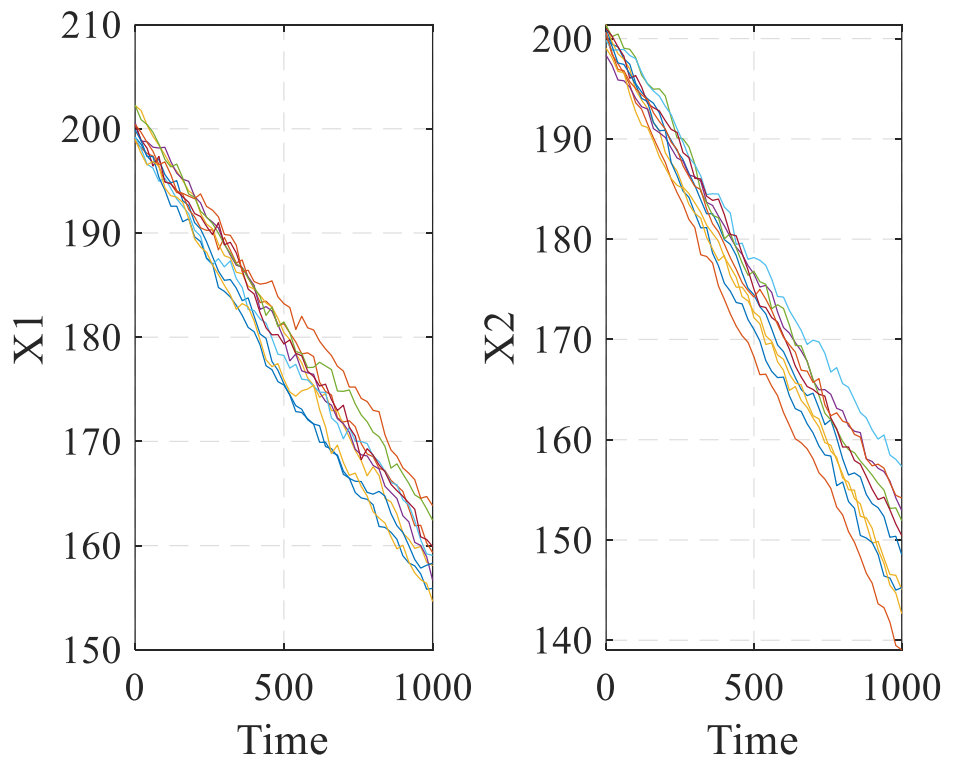

Fig. 3 An example of independent degradation paths with ten sample size.

### 3.3.2 Results

Fig. 4 illustrates the EMR of different causal discovery methods under two causal discovery strategies for independent degradation paths. In this case, $G$ in Eq. (7) is a 2×2 zero matrix. It can be seen that:

- When applying strategy $S_1$, all causal discovery techniques identify the independent degradation paths as causally related. This reveals the drawback of directly utilizing raw degradation data for causal discovery, as such data significantly violate the IID assumption required by causal discovery techniques.
- Under the proposed strategy $S_2$, Direct-LiNGAM, NOTEARS-Linear and NOTEARS-MLP can effectively distinguish two independent degradation processes as the sample size increases. In contrast, Stable-PC and GES have a risk of misjudgment regardless of sample size, and CaPS fails to identify independent degradation processes.
- Under strategy $S_2$, NOTEARS-Linear needs the smallest sample size for the robust

identification of independent relationships ($n \geq 2$), followed by NOTEARS-MLP ($n \geq 4$) and Direct-LiNGAM ($n = 8$).

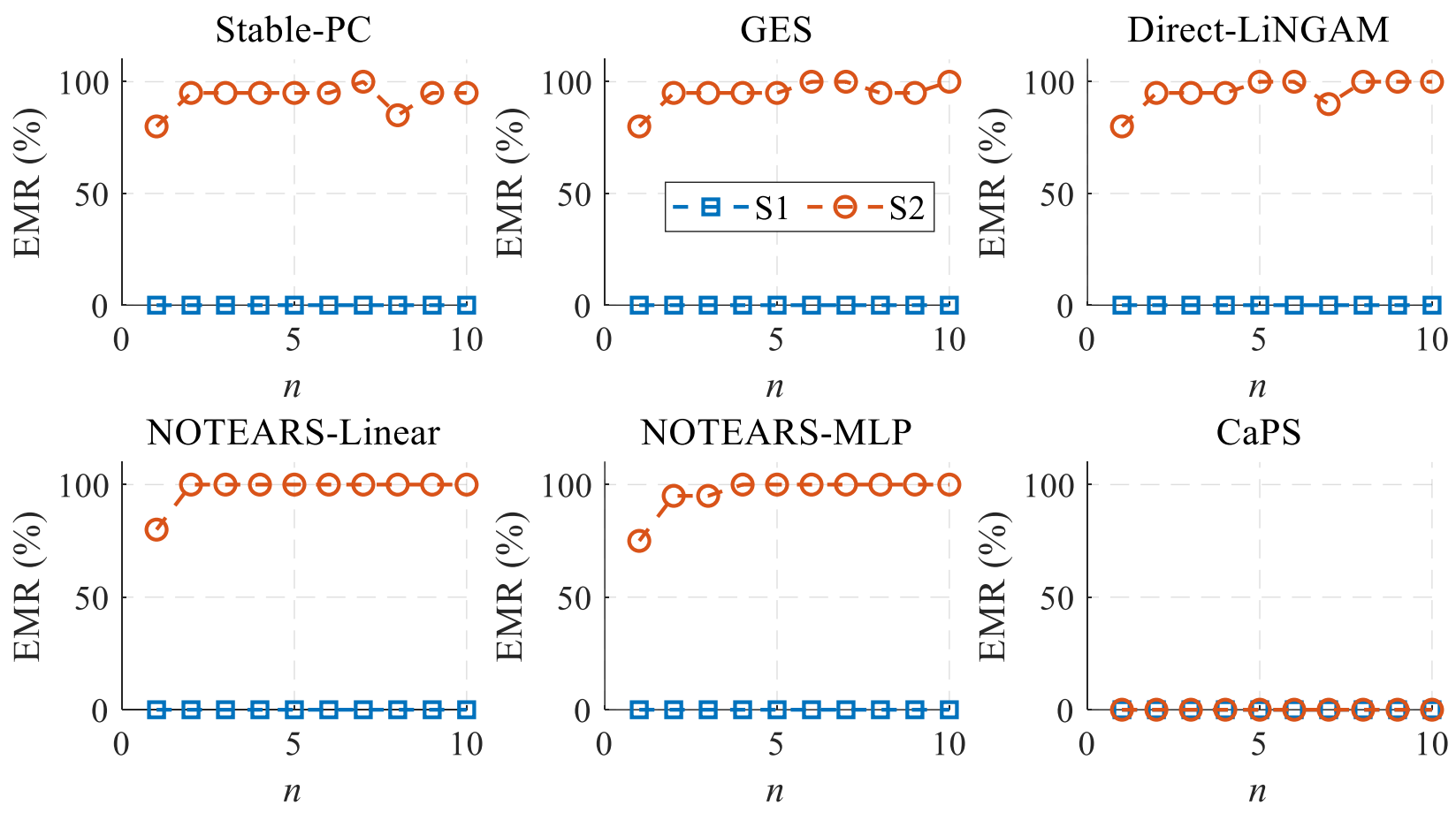

Fig. 4 EMR of different causal discovery methods under two causal discovery strategies for independent degradation paths.

### 3.3.3 Sensitivity analysis

In this subsection, we explore the effects of degradation nonlinearity ($\gamma$), the magnitude of random effects ($v_a$), the uncertainty magnitude in the time dimension ($\sigma$) and the magnitude of measurement error ($\sigma_\varepsilon$) on results of different causal discovery techniques through sensitivity analysis.

**Impact of degradation nonlinearity $\gamma$**: To examine the effect of $\gamma$, we vary its value as shown in Table 3. Notably, to ensure noticeable performance degradation, the values of $\mu$ are adjusted accordingly. The results are presented in Fig. 5. It can be seen that the accuracy of causal discovery is negatively affected as $\gamma$ decreases. When $\gamma$ falls below 1, the EMRs of Stable-PC, GES, and Direct-LiNGAM decline sharply. While NOTEARS-Linear and NOTEARS-MLP maintain stable results within this range, their EMRs drop significantly when $\gamma \leq 0.5$. Among them, NOTEARS-Linear demonstrates the best performance, accurately distinguishing independent degradation paths for $\gamma$ values between 0.75 and 1.5 when $n \geq 3$, followed by NOTEARS-MLP when $n \geq 9$.

Table 3 Parameter settings for assessing the impact of degradation nonlinearity on independent degradation paths.

| Test number | 1 | 2 | 3 | 4 | 5 |
|---|---|---|---|---|---|
| $\gamma$ | 0.5 | 0.75 | 1 | 1.25 | 1.5 |
| $\mu_{a_1}$ | -0.8 | -0.2 | -0.04 | -0.008 | -0.0008 |
| $\mu_{a_2}$ | -1 | -0.25 | -0.05 | -0.01 | -0.001 |

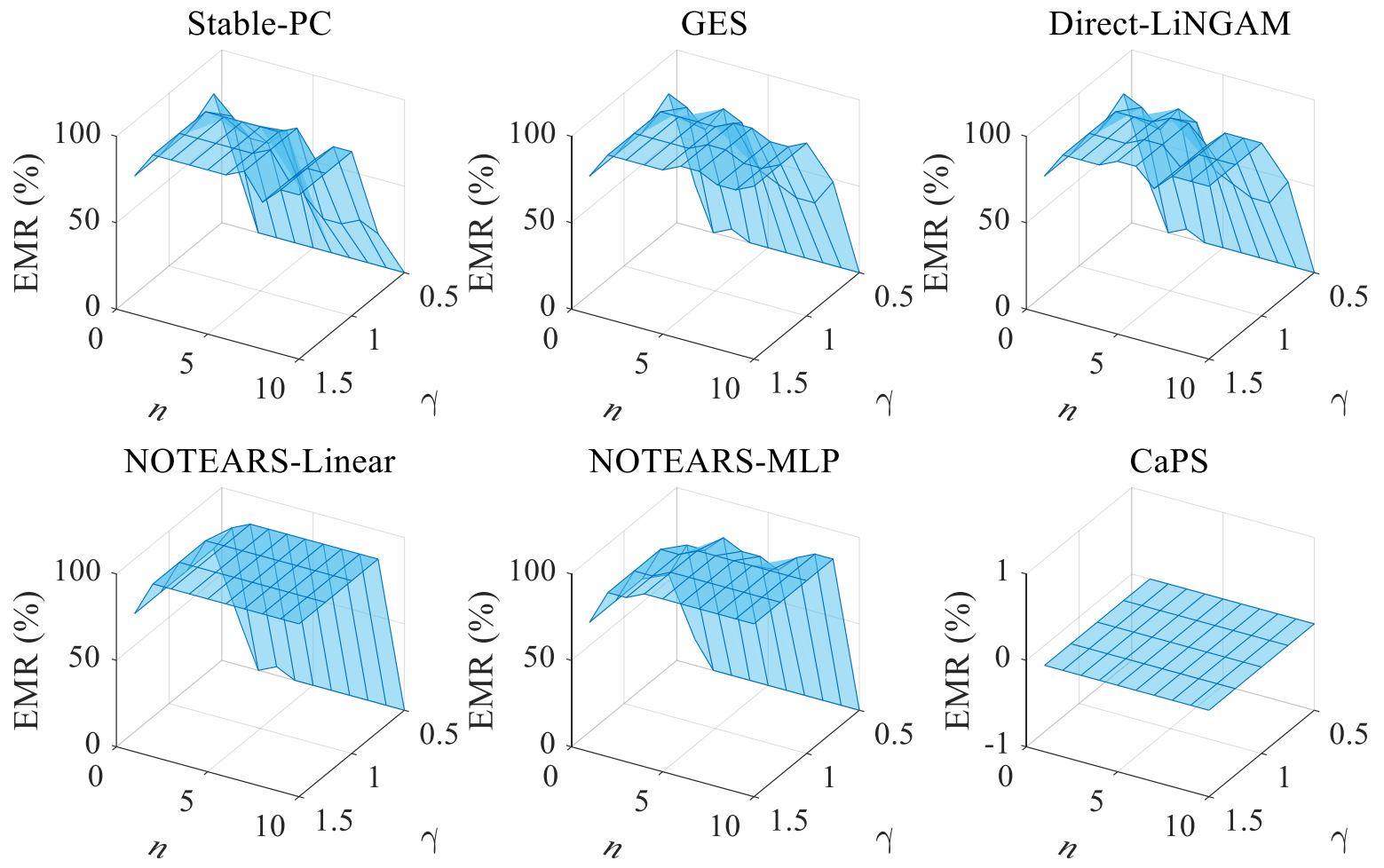

Fig. 5 EMR of different causal discovery techniques on independent degradation paths with varying sample sizes and degradation nonlinearity.

**Impact of random effect $\nu_a$**: To examine the effect of $\nu_a$, its value is varied among 0, 0.025, 0.05, 0.1 and 0.2, with the results illustrated in Fig. 6. It can be observed that the magnitude of random effect has a negligible effect on the accuracy of causal discovery. Among them, NOTEARS-Linear demonstrates the best performance, accurately distinguishing independent degradation paths for $\nu_a$ values between 0 and 0.2 when $n \geq 2$, followed by NOTEARS-MLP when $n \geq 4$.

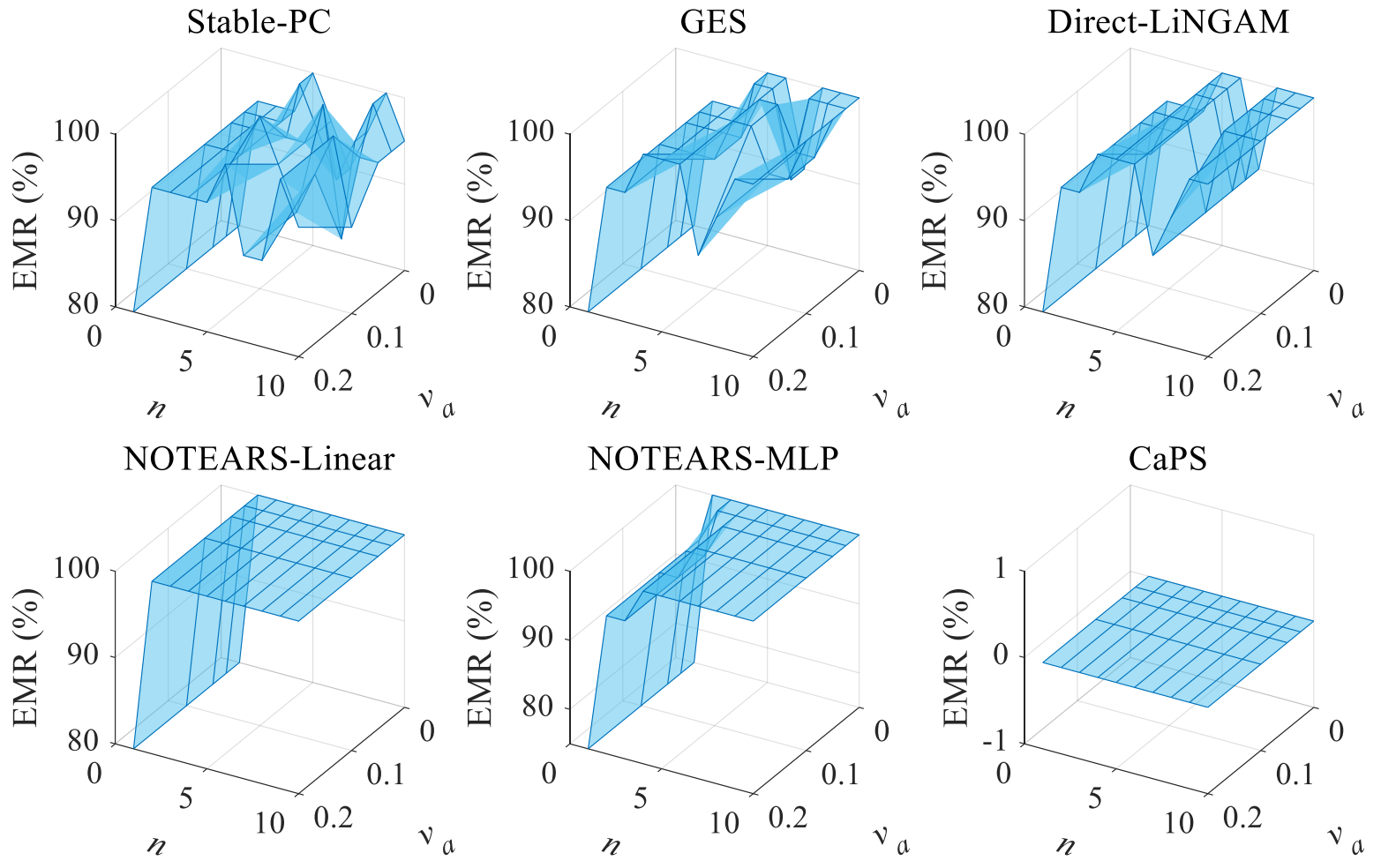

Fig. 6 EMR of different causal discovery techniques on independent degradation paths with varying sample sizes and random effects.

**Impact of measurement error $\sigma_\varepsilon$**: The impact of $\sigma_\varepsilon$ is examined by varying its value among 0, 0.2, 0.4, 0.8 and 1.6, with the results depicted in Fig. 7. The findings indicate that the magnitude of the measurement error has little effect on causal discovery accuracy. Among all the methods, NOTEARS-Linear achieves the highest performance, correctly identifying independent degradation paths across $\sigma_\varepsilon$ values from 0 to 1.6 when $n \geq 2$, with NOTEARS-MLP following when $n \geq 4$.

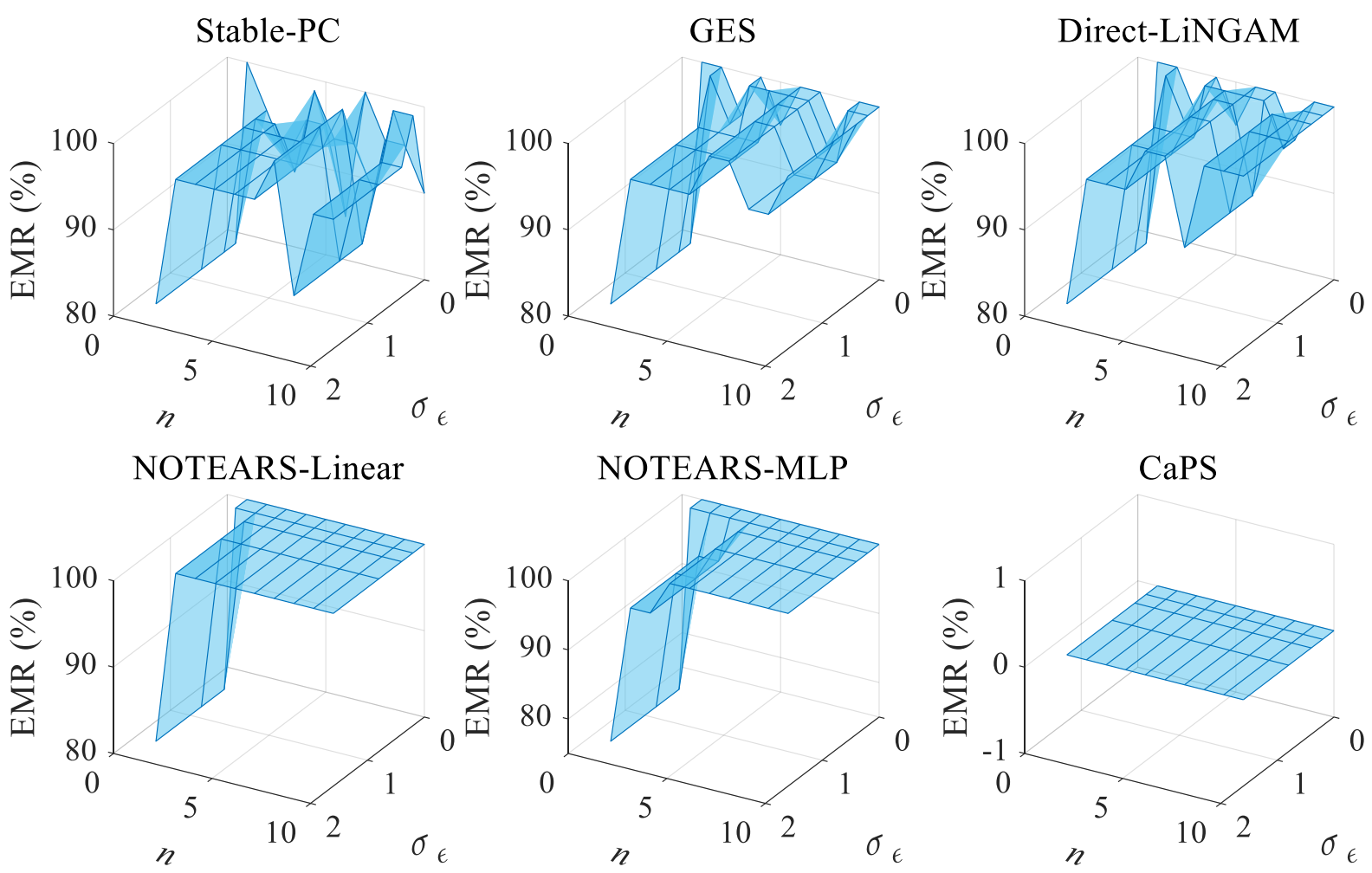

Fig. 7 EMR of different causal discovery techniques on independent degradation paths with varying sample sizes and measurement errors.

**Impact of diffusion coefficient $\sigma$**: The impact of $\sigma$ is investigated by adjusting its value across 0, 0.05, 0.1, 0.2 and 0.4, with the results shown in Fig. 8. The results indicate that the magnitude of diffusion coefficient has a negative impact on causal discovery accuracy, especially for NOTEARS-MLP. Among all methods, NOTEARS-Linear exhibits the highest performance, successfully distinguishing independent degradation paths for $\sigma$ between 0 and 1.6 when $n \geq 4$, followed by NOTEARS-MLP when $n \geq 9$.

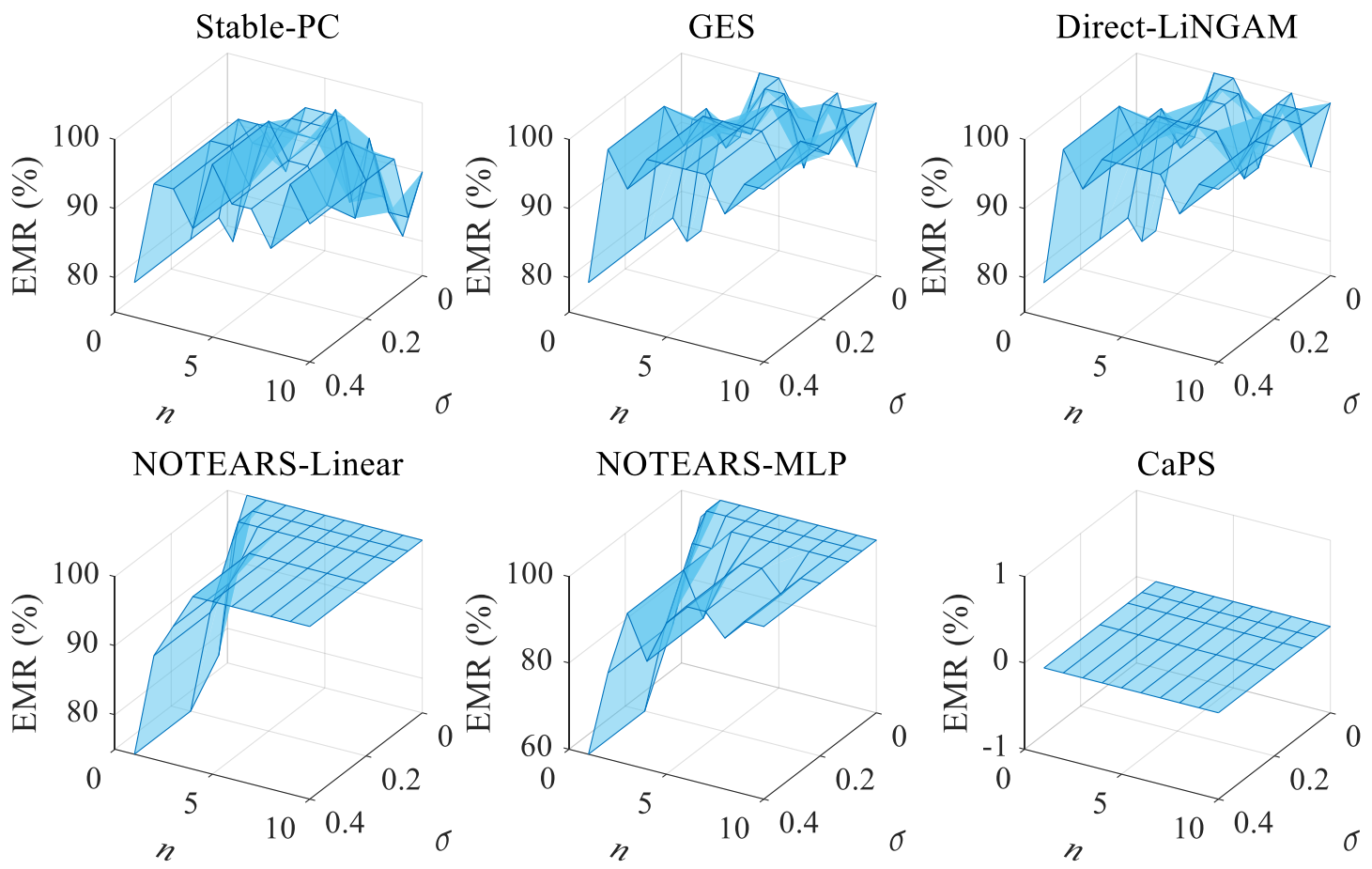

Fig. 8 EMR of different causal discovery techniques on independent degradation paths with varying sample sizes and diffusion coefficients.

## 3.4 Causal discovery between causally dependent degradation paths

### 3.4.1 Numerical experiment design

For causally dependent degradation paths, the parameter settings of Eq. (5) are almost the same as in Section 3.3.1, and are listed in Table 4. The difference is that the initial performance of $X_2$ is zero and a causal relationship from $X_1$ to $X_2$ exists, as described by $g_2\left(X_1\left(t\right)\right) = \alpha X_1\left(t\right)^{\beta}$.

Table 4 Parameter settings of causally dependent degradation paths

| Parameters | Values | Parameters | Values |
|---|---|---|---|
| $\mu_{X_{10}}$ | 200 | $\mu_{X_{20}}$ | 0 |
| $\sigma_{X_{10}}$ | 1 | $\sigma_{X_{20}}$ | 0 |
| $\mu_{a_1}$ | -0.04 | $\mu_{a_2}$ | -0.05 |
| $v_{a_1}$ | 0.05 | $v_{a_2}$ | 0.05 |
| $\sigma_1$ | 0.1 | $\sigma_2$ | 0.1 |
| $\sigma_{\varepsilon_1}$ | 0.4 | $\sigma_{\varepsilon_2}$ | 0.4 |
| $\gamma_1$ | 1 | $\gamma_2$ | 1 |
| $\Delta t$ | 20 | $m$ | 51 |
| $\alpha$ | 1 | $\beta$ | 1 |

Similarly, the sample size ranges from one to ten, and twenty sets of degradation paths are simulated to evaluate robustness ($N$ = 20). An example of causally dependent degradation paths with ten sample size is shown in Fig. 9.

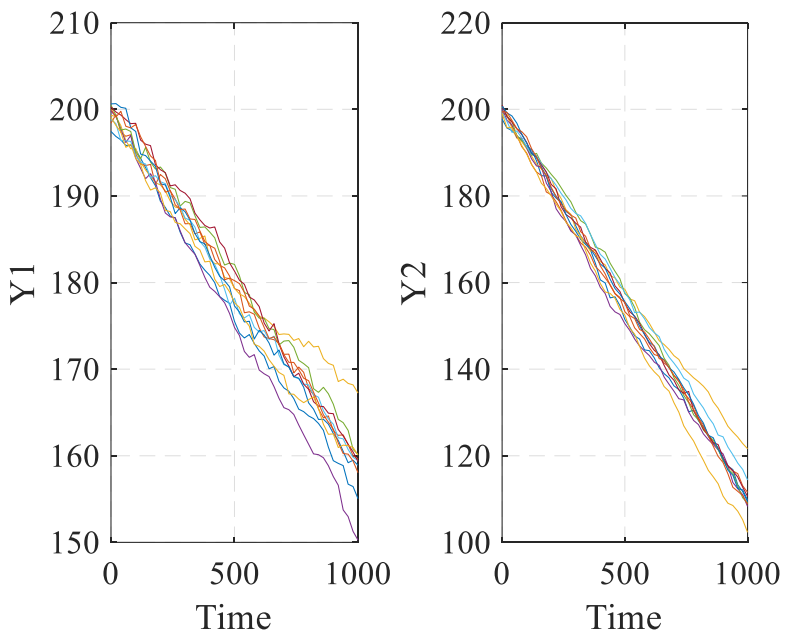

Fig. 9 An example of causally dependent degradation paths with ten sample size.

3.4.2 Results

Fig. 10 presents the EMR of different causal discovery methods under two causal discovery strategies for causally dependent degradation paths. In this case, $G$ in Eq. (7) is [0 1; 0 0], which represents a causal link from $X_1$ to $X_2$. It can be seen that:

- When applying strategy $S_1$, NOTEARS-Linear, NOTEARS-MLP and CaPS consistently detect the correct causal relationships, whereas Direct-LiNGAM identifies them randomly, and Stable-PC and GES fail entirely across sample sizes. However, when considered alongside the results in Fig. 4, NOTEARS-Linear, NOTEARS-MLP and CaPS in fact indiscriminately identify a causal relationship from $X_1$ to $X_2$, regardless of whether the degradation paths are independent or causally dependent. This demonstrates the flaw of using the strategy $S_1$.
- Under the proposed strategy $S_2$, Stable-PC, GES and Direct-LiNGAM yield results nearly identical to those under strategy S1, failing to accurately capture causal relationships. In contrast, NOTEARS-Linear, NOTEARS-MLP and CaPS effectively identify causal relationships as the sample size increases. However, when considered alongside the results in Fig. 4, CaPS fails to distinguish independent degradation paths, whereas NOTEARS-Linear and NOTEARS-MLP exhibit satisfactory performance in both independent and causally dependent degradation paths.

- Under strategy $S_2$, only NOTEARS-MLP provides robust identification of causal relationships ($n \geq 7$).

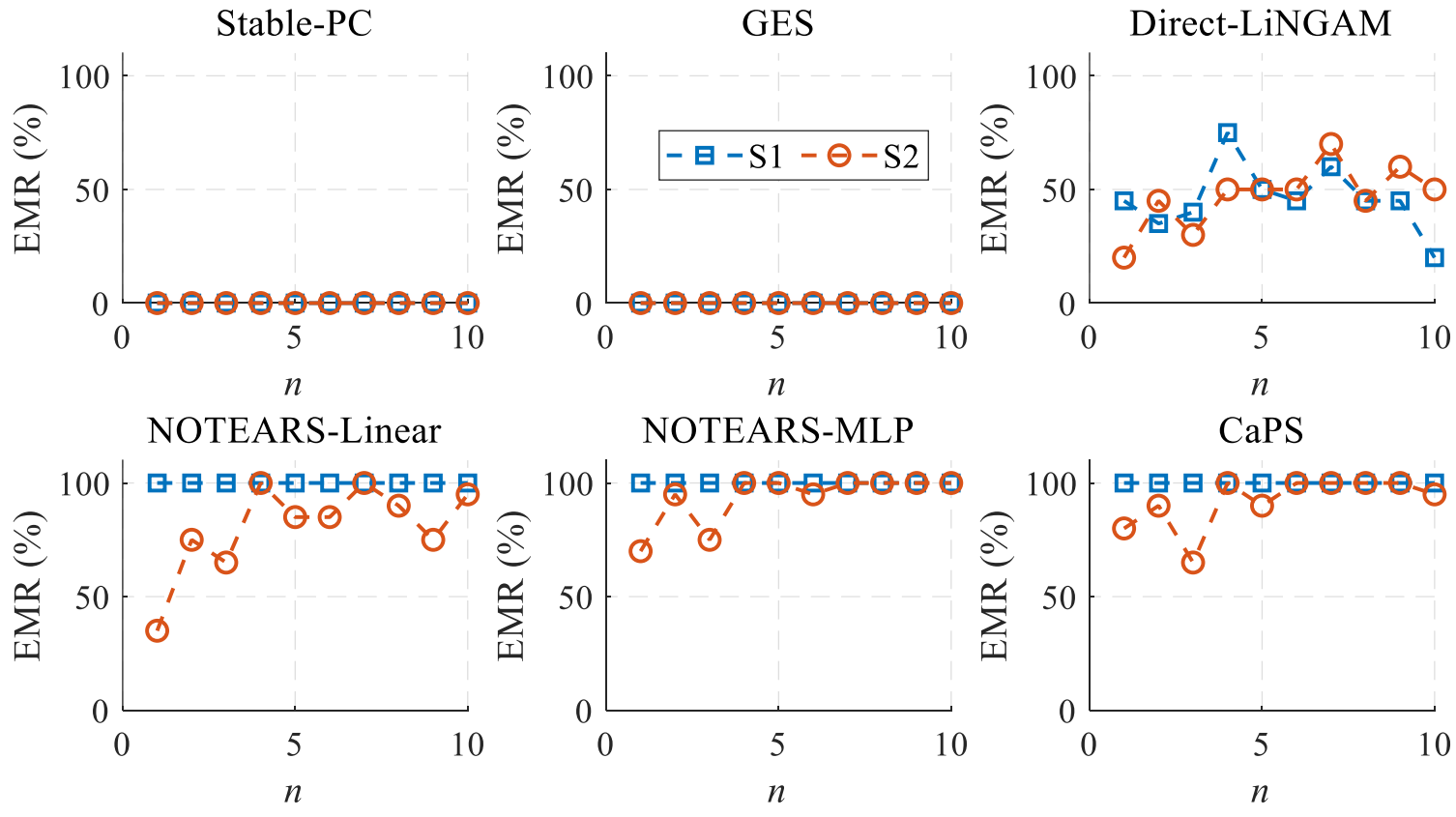

Fig. 10 EMR of different causal discovery methods under two causal discovery strategies for independent degradation paths.

To enhance the understanding of results obtained by different causal discovery techniques, Fig. 11 illustrates how the result frequency of various methods changes under strategy $S_2$ as the sample size increases. While Stable-PC and GES can detect the existence of causal relationships when $n \geq 4$, they fail to infer causal direction. This limitation arises because the target causal relationship cannot be determined through a V-structure, leading them to provide MECs without causal direction. Similarly, Direct-LiNGAM can identify causal relationships but fails to determine their direction. In contrast, NOTEARS-Linear, NOTEARS-MLP and CaPS accurately infer causal direction, with NOTEARS-MLP providing robust identification when $n \geq 7$.

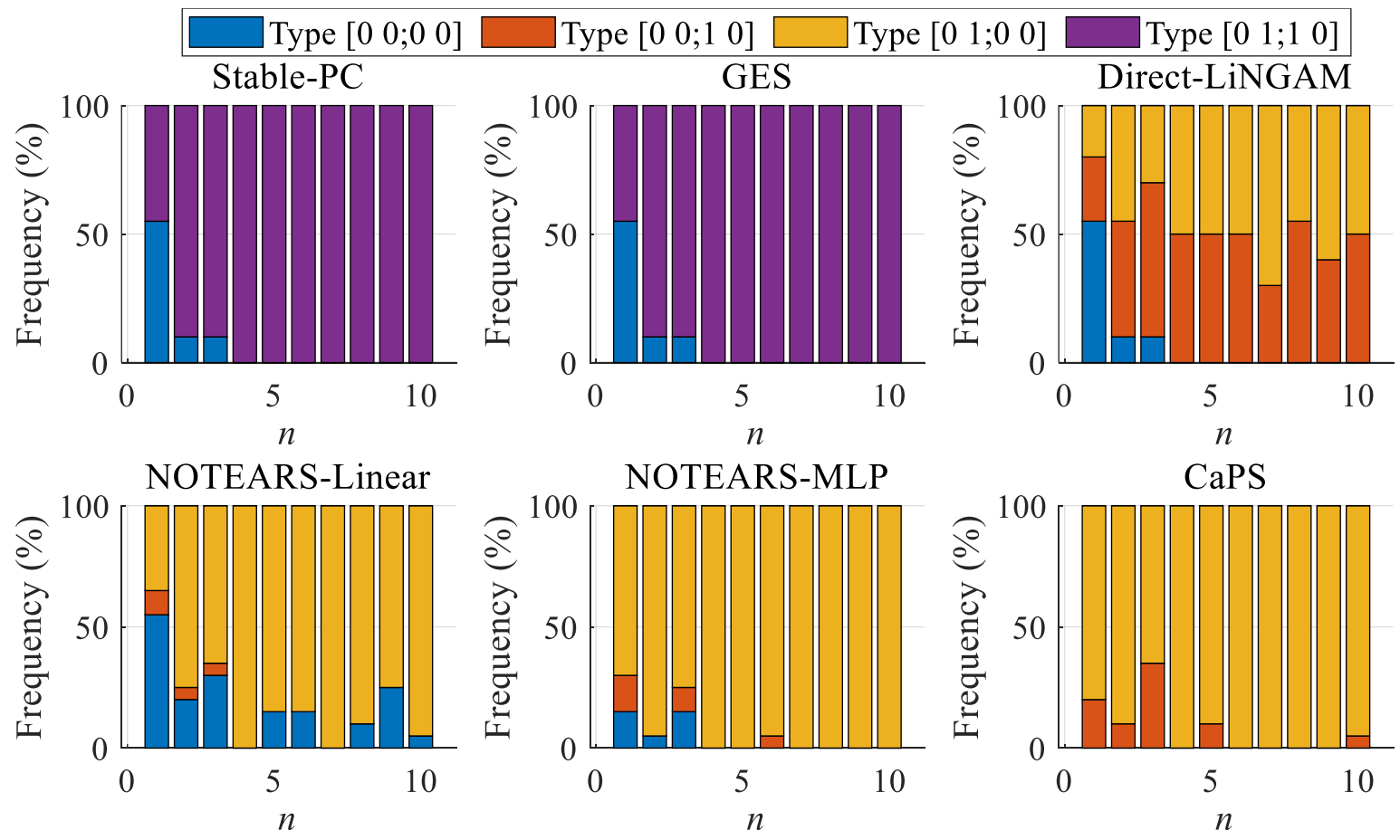

Fig. 11 Result frequency of different causal discovery techniques on causally dependent degradation paths under strategy $S_2$ with varying sample sizes.

### 3.4.3 Sensitivity analysis

In this subsection, we explore the effects of causal relationship nonlinearity ($\beta$), degradation nonlinearity ($\gamma$), magnitude of random effects ($v_a$), uncertainty magnitude in the time dimension ($\sigma$) and magnitude of measurement error ($\sigma_\varepsilon$) on the results of different causal discovery techniques through sensitivity analysis.

**Impact of the causality nonlinearity factor $\boldsymbol{\beta}$**: To examine the effect of $\beta$, we vary its value as listed in Table 5. Notably, to ensure noticeable performance degradation, the values of $\alpha$ are adjusted accordingly. Fig. 12 displays the degradation paths of $X_2$ with boundaries at 95% confidence level across different values of $\beta$. The EMR results are presented in Fig. 13. It can be seen that the accuracy of causal discovery decreases as $\beta$ decreases. When $\beta \leq 0.8$, none of the methods can robustly identify causal relationships. NOTEARS-MLP can stably determine causal direction when $\beta \geq 1$ ($n \geq 7$), whereas NOTEARS-Linear and CaPS require $\beta \geq 1.2$ to achieve stable causal direction identification ($n \geq 7$ and $n \geq 4$). The result frequencies are presented in the supplementary material.

Table 5 Parameter settings for assessing the impact of causality nonlinearity on causally dependent degradation paths.

| Test number | 1 | 2 | 3 | 4 | 5 | 6 | 7 | 8 | 9 |
|---|---|---|---|---|---|---|---|---|---|
| $\beta$ | 0.2 | 0.4 | 0.6 | 0.8 | 1 | 1.2 | 1.4 | 1.6 | 1.8 |
| $\alpha$ | 70 | 24 | 8.4 | 2.9 | 1 | 0.345 | 0.119 | 0.042 | 0.014 |

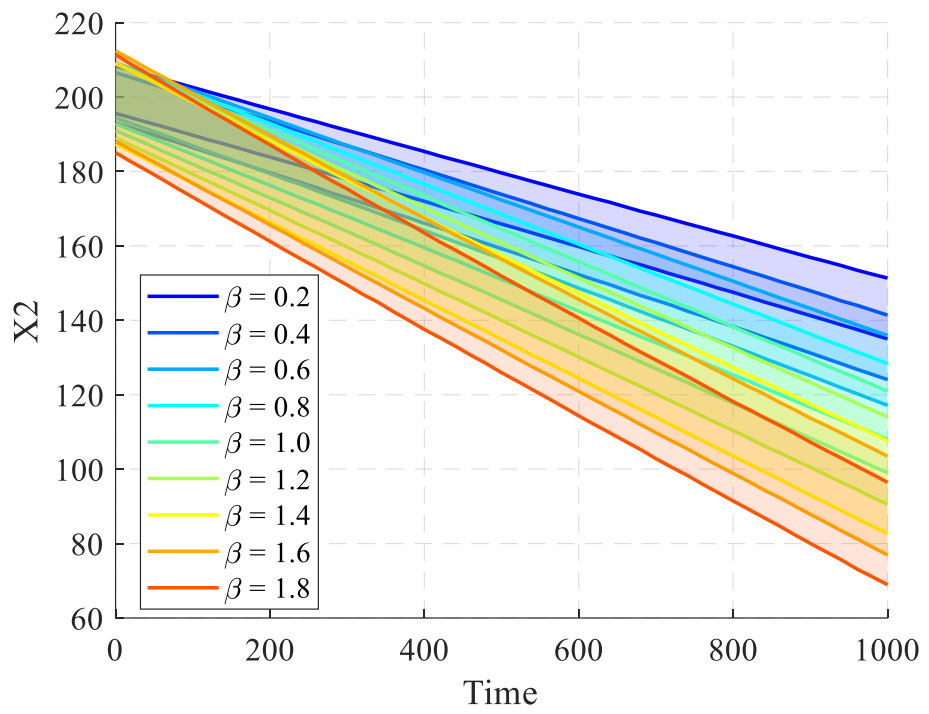

Fig. 12 Degradation paths of $X_2$ with boundaries at 95% confidence level across different values of $\beta$.

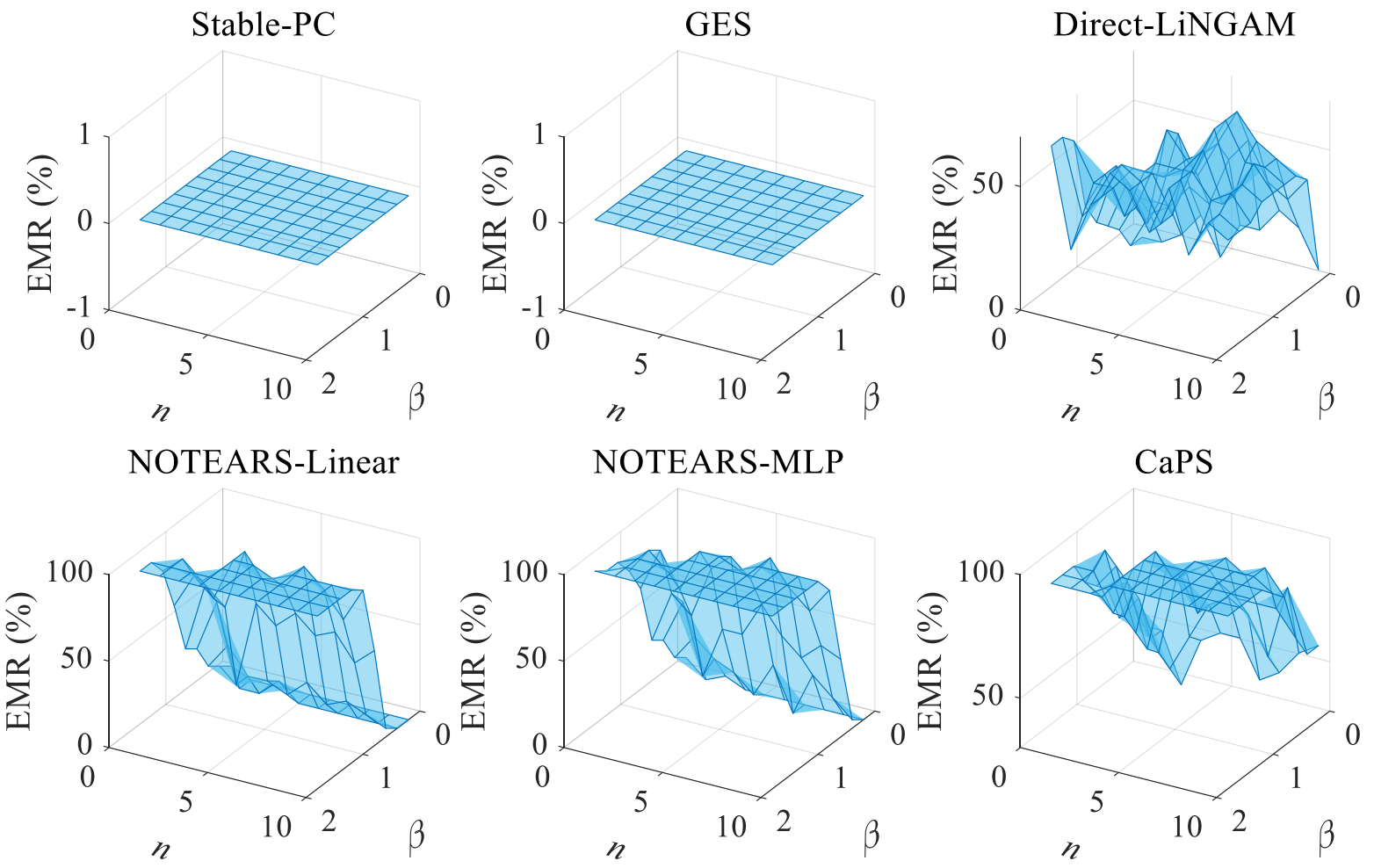

Fig. 13 EMR of different causal discovery techniques on causally dependent degradation paths with varying sample sizes and causality nonlinearity.

**Impact of degradation nonlinearity $\gamma$**: To examine the effect of $\gamma$, we vary its value as listed in Table 3. The EMR results are shown in Fig. 14, and the result frequencies are presented in the supplementary material. The findings indicate that causal discovery accuracy is lowest when the degradation process is linear. In this case, NOTEARS-Linear and CaPS have a risk of misjudgment in causal identification. Among all methods, NOTEARS-MLP demonstrates superior performance, accurately distinguishing causally dependent degradation paths for $\gamma$ values between 0.5 and 1.5 when $n \geq 7$.

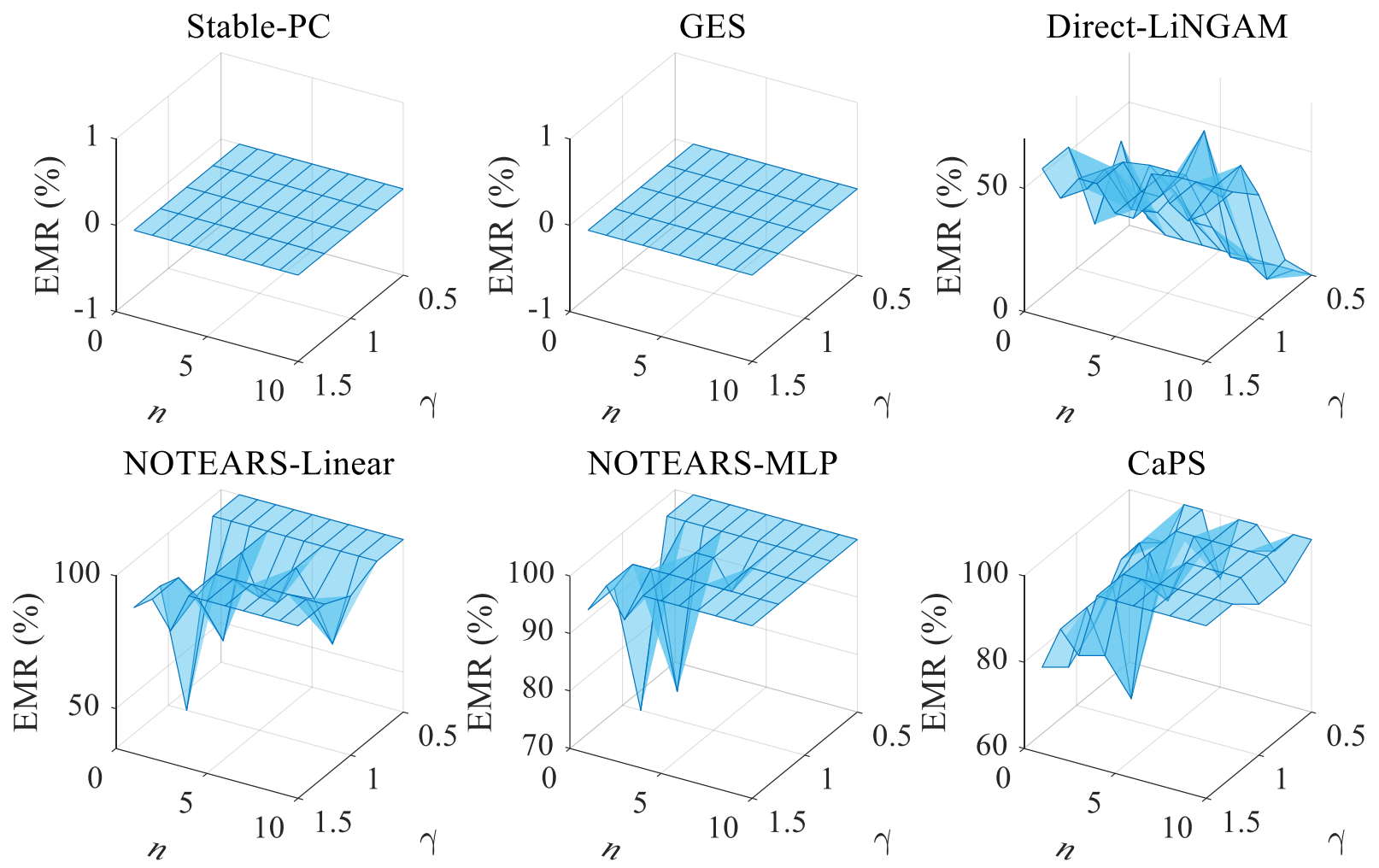

Fig. 14 EMR of different causal discovery techniques on independent degradation paths with varying sample sizes and degradation nonlinearity.

**Impact of random effect $v_a$**: To examine the effect of $v_a$, its value is varied among 0, 0.025, 0.05, 0.1 and 0.2, with the results illustrated in Fig. 15. It can be observed that, in the range considered, the magnitude of random effect has a negligible effect on the accuracy of causal discovery. Among them, NOTEARS-MLP demonstrates the best performance, accurately distinguishing causally dependent degradation paths for $v_a$ values between 0 and 0.2 when $n \geq 7$.

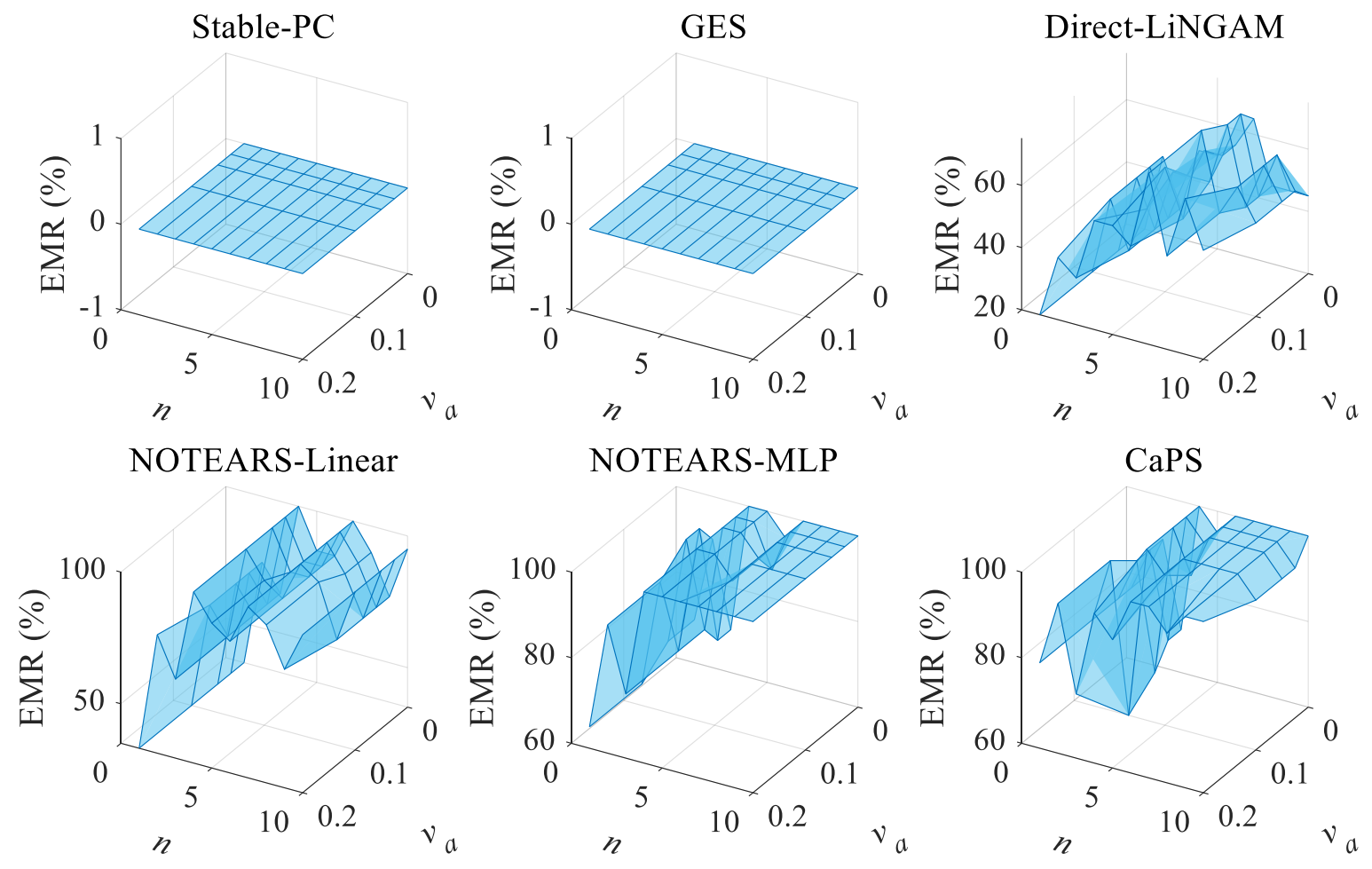

Fig. 15 EMR of different causal discovery techniques on independent degradation paths with varying sample sizes and random effects.

**Impact of measurement error $\sigma_\varepsilon$**: The impact of $\sigma_\varepsilon$ is examined by varying its value among 0, 0.2, 0.4, 0.8 and 1.6, with the results depicted in Fig. 7. The findings indicate that the magnitude of the measurement error has significant effect on causal discovery accuracy. When $\sigma_\varepsilon \geq 0.8$, none of the techniques can correctly provide a robust result. Among all the methods, NOTEARS-MLP achieves the highest performance, correctly identifying causally dependent degradation paths across $\sigma_\varepsilon$ values from 0 to 0.4 when $n \geq 7$.

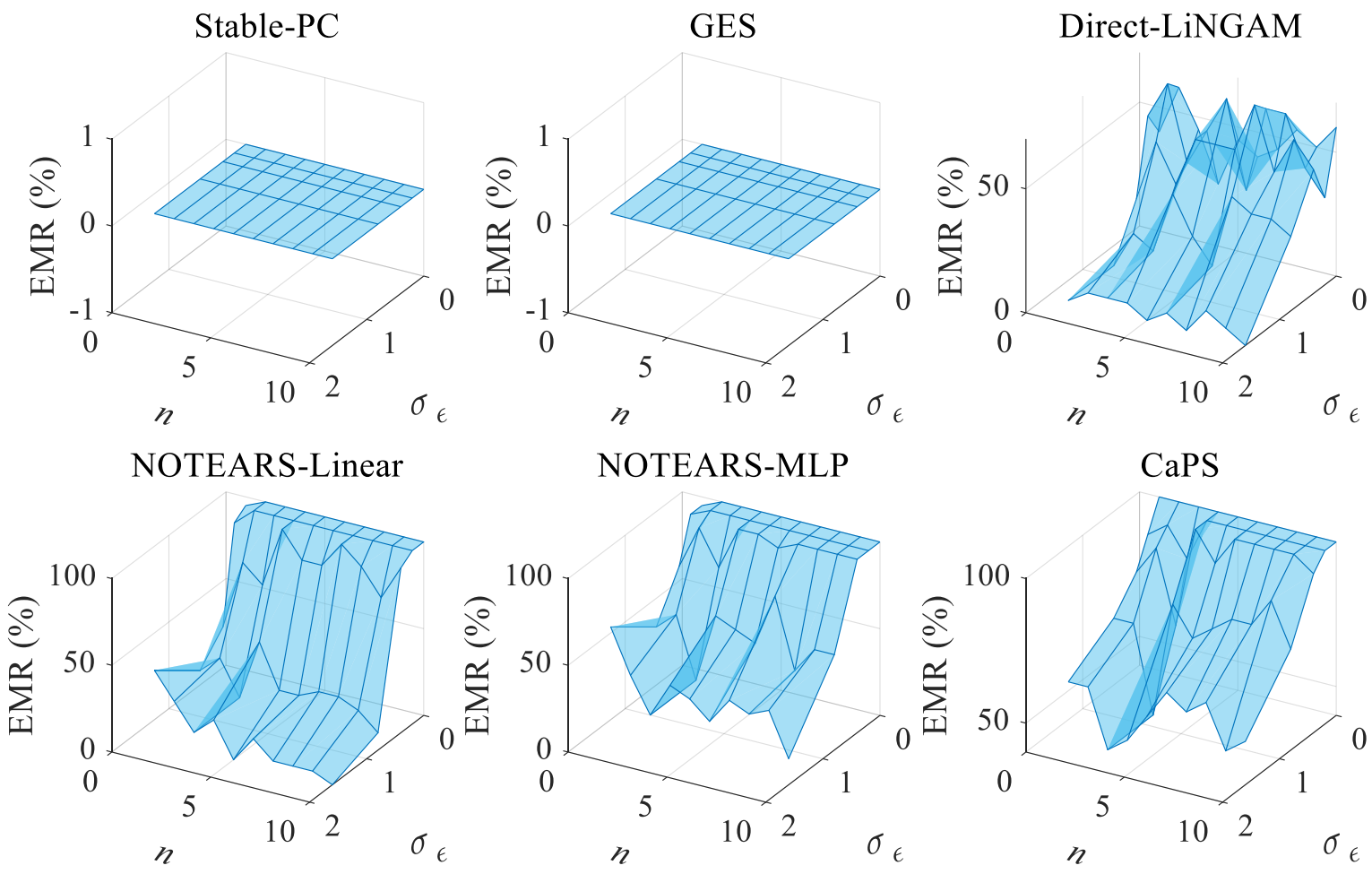

Fig. 16 EMR of different causal discovery techniques on independent degradation paths with varying sample sizes and measurement errors.

**Impact of diffusion coefficient *σ***: The impact of $\sigma$ is investigated by adjusting its value across 0, 0.05, 0.1, 0.2 and 0.4, with the results shown in Fig. 8. Surprisingly, the results indicate that the magnitude of the diffusion coefficient has a positive impact on causal discovery accuracy. When $\sigma \leq 0.05$, none of the techniques can correctly provide robust results. Among all methods, NOTEARS-MLP exhibits the highest performance, successfully distinguishing causally dependent degradation paths for $\sigma$ between 0.1 and 0.4 when $n \geq 7$.

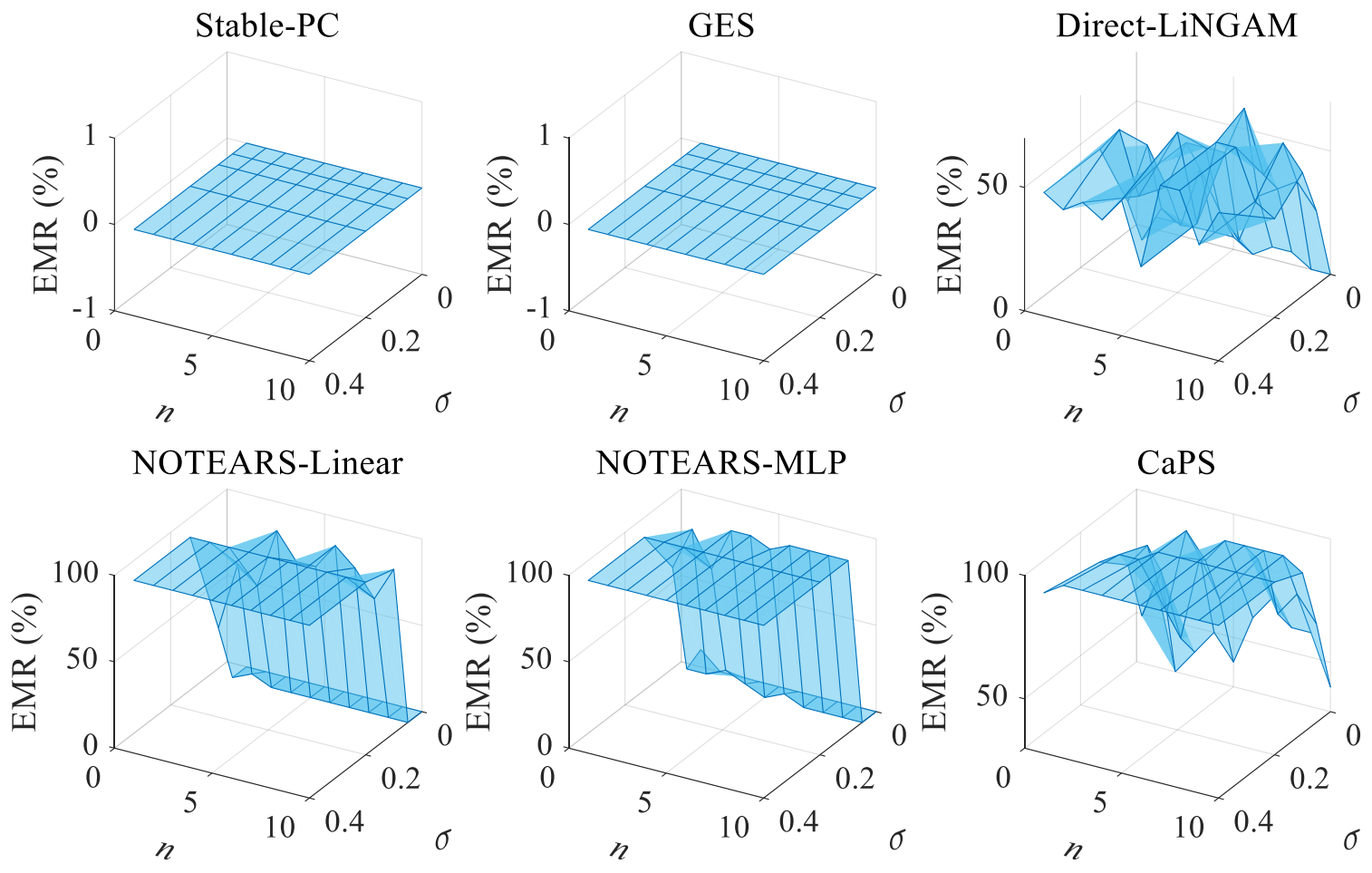

Fig. 17 EMR of different causal discovery techniques on independent degradation paths with varying sample sizes and diffusion coefficients.

## 3.5 Discussions

### 3.5.1 Results of temporal causal discovery techniques

To demonstrate that temporal causal discovery techniques are unsuitable for degradation data recorded in a steady state, we employ the numerical degradation data at $n$ = 1, and compare the causal discovery results obtained by GC [53], TE [54] and CCM [55]. For temporal techniques, causal discovery strategy $S_1$ is naturally applied.

From Fig. 18, for independent degradation paths, GC detects a moderate bidirectional causal relationship, TE suggests no causality, and CCM indicates a strong bidirectional causal relationship. Then, as illustrated in Fig. 19, for degradation paths with a causal link from $X_1$ to $X_2$, GC incorrectly suggests a causal influence of $X_2$ on $X_1$, TE again reports no causality, and CCM continues to indicate a bidirectional causal relationship. These findings indicate that temporal causal discovery techniques fail to identify the true dependent causal relationships in degradation data.

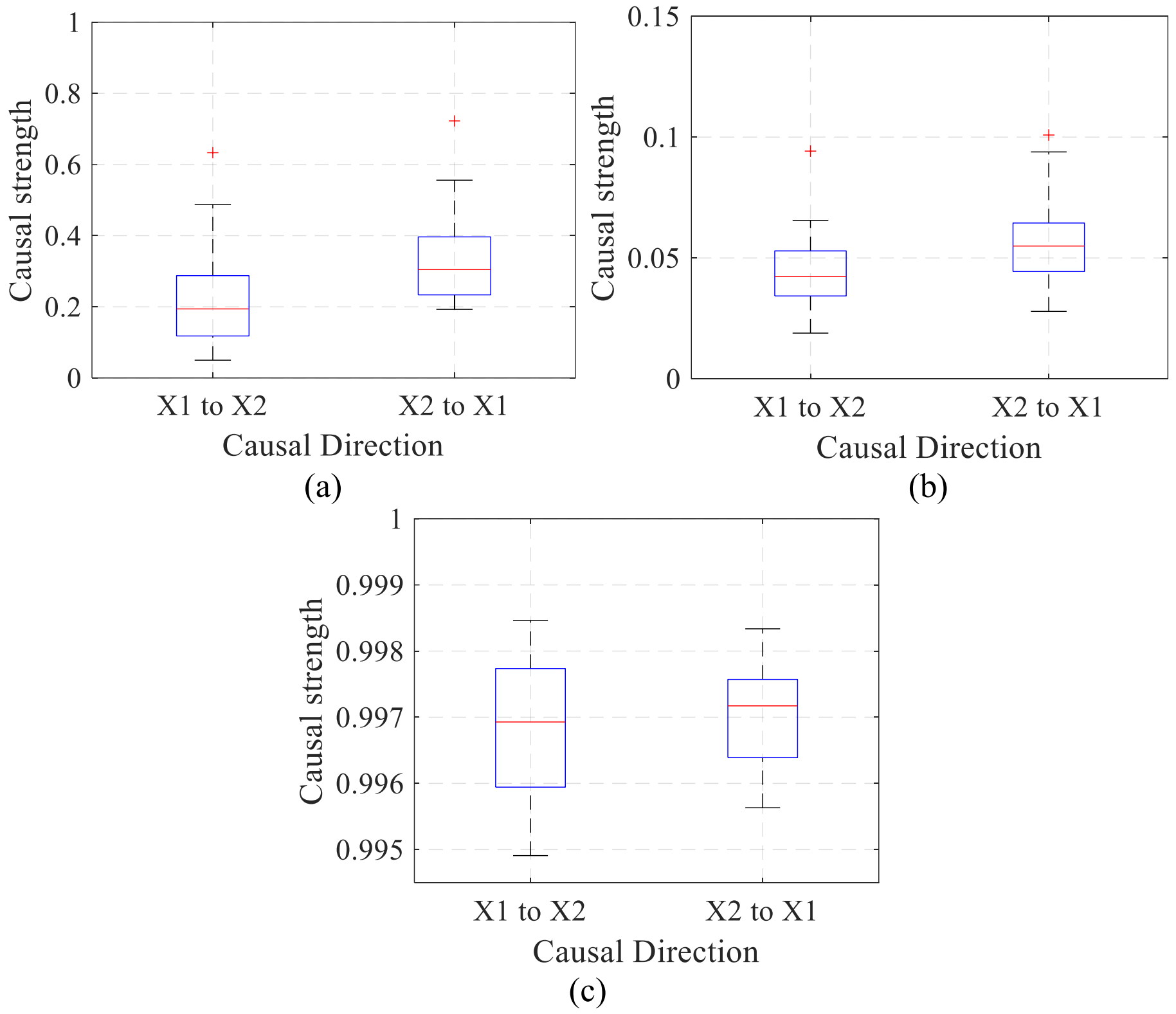

Fig. 18 Causal strength of different temporal causal discovery techniques on independent degradation paths (a) GC (b) TE (c) CCM.

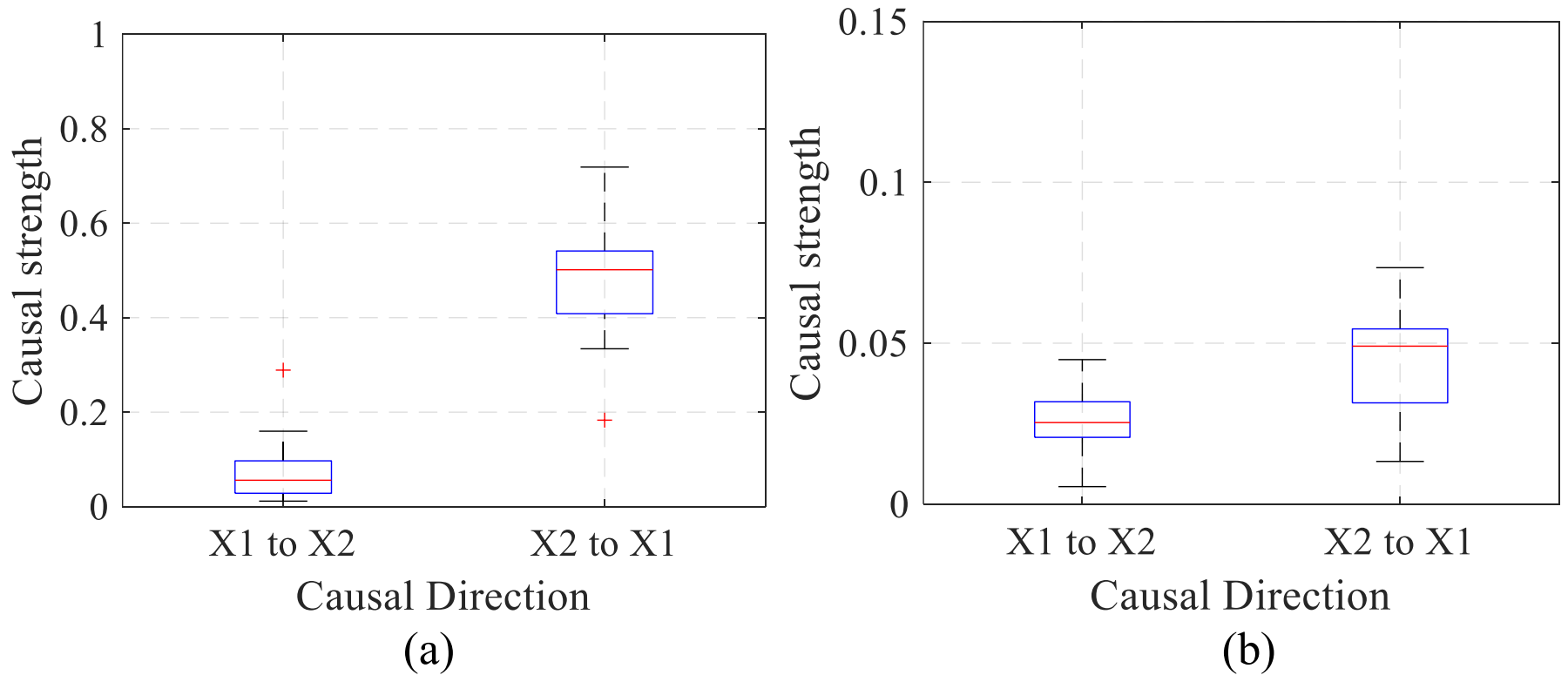

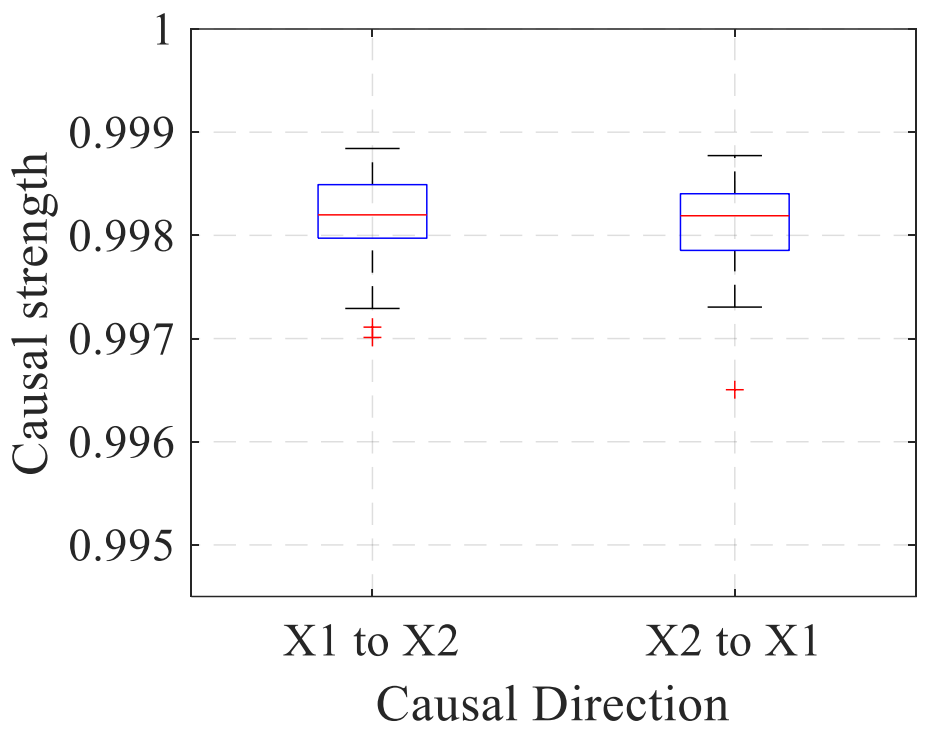

(c)

Fig. 19 Causal strength of different temporal causal discovery techniques on causally dependent degradation paths (a) GC (b) TE (c) CCM.

### 3.5.2 Summary on the impacts of various factors on causal discovery

Based on the sensitivity analysis in Section 3, we can deduce the impacts of various factors on causal discovery, which is summarized in Table 6 and described as follows:

- Causality nonlinearity ($\beta$): The accuracy of causal discovery decreases as $\beta$ decreases, which indicates that current techniques are incapable of detecting weak causal relationships between degradation paths. When $\beta \leq 0.8$, none of the methods can robustly identify causal relationships. But for $\beta \geq 1$, NOTEARS-MLP can stably determine causal relationships when $n \geq 7$.
- Degradation nonlinearity ($\gamma$): The accuracy of independent degradation identification is negatively affected as $\gamma$ decreases. When $\gamma \leq 0.5$, none of the techniques can present correctly robust results, but for $\gamma \geq 0.75$, NOTEARS-Linear and NOTEARS-MLP can give robust results when $n \geq 3$ and $n \geq 9$, respectively. For causally dependent degradation paths, causal discovery accuracy is lowest when the degradation process is linear, but NOTEARS-MLP is still capable of accurately distinguishing causal relationships across different values of $\gamma$.
- Random effect ($v_a$): The magnitude of random effect has a negligible effect on the accuracy of causal discovery, whether for independent or causally dependent degradation paths.
- Measurement error ($\sigma_\varepsilon$): The magnitude of the measurement error has little effect on the accuracy of independent degradation identification. However, for causally dependent degradation paths, it has a significant impact. When $\sigma_\varepsilon \geq 0.8$, none of the techniques can correctly provide robust results, but for $\sigma_\varepsilon \leq 0.4$, NOTEARS-MLP can give stably correct causal discovery results when $n \geq 7$.
- Diffusion coefficient ($\sigma$): The magnitude of diffusion coefficient has a negative impact on independent degradation identification, but NOTEARS-Linear and NOTEARS-MLP are still capable of accurately distinguishing it across different values of $\sigma$. However, for causally dependent degradation paths, the magnitude of the diffusion coefficient has a positive impact. When $\sigma \leq 0.05$, none of the techniques can correctly provide robust result, but for $\sigma \geq 0.1$, NOTEARS-MLP successfully distinguishes causal relationships when $n \geq 7$.

Consequently, to reliably identify both independent and causally dependent degradation paths, the following conditions should be met: $\beta \geq 1$, $\gamma \geq 0.75$, $\sigma_\varepsilon \leq 0.4$, $\sigma \geq 0.1$, and the appropriate method to use is NOTEARS-MLP with a sample size of $n \geq 9$. Although the constrains on these parameters may vary depending on the values of other parameters in Tables 2 and 3, it still provides guidance for assessing the feasibility of causal discovery between degradation paths.

Table 6 Impacts of various factors on causal discovery in the numerical study.

| Methods | Independent degradation | Causally dependent degradation |
|---|---|---|
| Causality nonlinearity | / | Positive and significant impact. When $\beta \leq 0.8$, no methods are effective |
| Degradation nonlinearity | Positive and significant impact. When $\gamma \leq 0.5$, no methods are effective | Moderate impact. The impact is greatest when $\gamma = 1$ |
| Random effect | Slight impact | Slight impact |
| Measurement error | Negative and moderate impact. | Negative and significant impact. When $\sigma_\varepsilon \geq 0.8$, no methods are effective |
| Diffusion coefficient | Negative and moderate impact. | Positive and significant impact. When $\sigma \leq 0.05$, no methods are effective |

3.5.3 Summary on the characteristics of different causal discovery techniques

Based on the results in Section 3, we can deduce the characteristic of each causal discovery technique, which is summarized in Table 7 and described as follows:

- Stable-PC and GES: The performance of these two methods is nearly identical. For independent degradation paths, both methods achieve a high success rate in independence identification, though there remains a small risk of failure. For causally dependent degradation paths, both methods can recognize the existence of causal relationships but fail to determine the causal direction, since the causal direction between two variables cannot be inferred using the V-structure. Additionally, as demonstrated in the supplementary material, both methods show a significant advantage in detecting the existence of causal relationships when $0.4 \leq \beta \leq 0.6$ compared to other techniques.
- Direct-LiNGAM: This method can distinguish independent degradation paths. However, it fails to identify the causally dependent degradation paths and randomly presents a causal direction.
- NOTEARS-Linear and NOTEARS-MLP: These two methods exhibit strong performance on both independent and causally dependent degradation paths. NOTEARS-Linear outperforms in identifying independent degradation paths due to its faster sample convergence. However, for causally dependent degradation paths, NOTEARS-MLP is more effective, delivering stable and correct results in most situations.
- CaPS: This method is incapable of detecting independent degradation paths. However, it can infer the direction of causal relationships when the existence of causal relationship is already confirmed. Besides, according to the supplementary material, in specific scenarios ($\beta = 0.6$), it achieves a marginally better performance than NOTEARS-MLP in determining causal direction.

Table 7  Characteristic of each causal discovery technique in the numerical study.

| Methods | Independent degradation | Causally dependent degradation |
|---|---|---|
| Stable-PC | Achieve a high success rate of independence identification, but remain a risk of failure | Achieve causal relationship identification, but fail to identify causal direction |
| GES | Achieve a high success rate of independence identification, but remain a risk of failure | Achieve causal relationship identification, but fail to identify causal direction |
| Direct-LiNGAM | Achieve a high success rate of independence identification, but remain a risk of failure | Fail to identify causal relationship |
| NOTEARS-Linear | Achieve a high success rate of independence identification | Achieve a high success rate, but remain a risk of failure |
| NOTEARS-MLP | Achieve a high success rate of independence identification, but inferior to NOTEARS-Linear in terms of sample convergence | Achieve a high success rate |
| CaPS | Fail to identify independence | Achieve a high success rate, but remain a risk of failure |

# 4 Engineering applications

## 4.1 A second-order multiple-feedback band pass filter

### 4.1.1 Dataset description

Second-order multiple-feedback band pass filters are canonical analog systems widely used in signal processing, communication systems and instrumentation [56]. The filter selectively amplifies signals within a specified frequency band while suppressing those outside it, making it particularly suitable for channel selection, noise reduction and harmonic isolation. A typical configuration includes three resistors ($R_1$, $R_2$, $R_3$), two capacitors ($C_1$, $C_2$) and a single operational amplifier, as illustrated in Fig. 20.

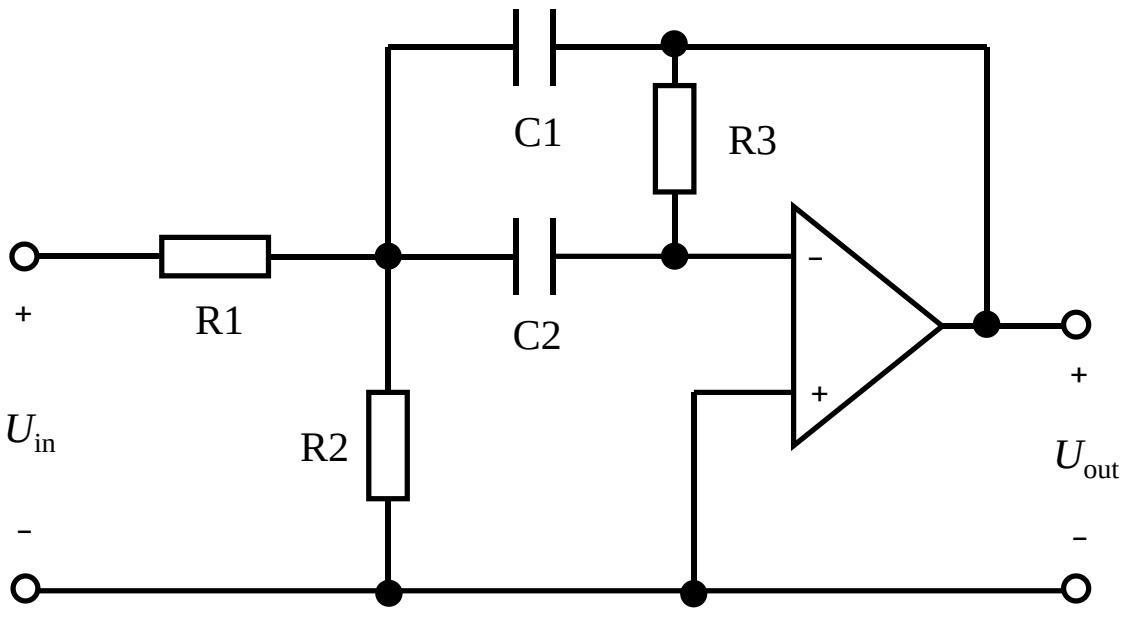

Fig. 20  Schematic of a second-order multiple feedback band pass filter.

In this study, we focus on two critical performance parameters of the filter: the center frequency $f_0$ and the peak gain $G$. The center frequency $f_0$ determines the pass band location and is essential for maintaining alignment with the target signal frequency. The peak gain $G$ represents the circuit's amplification at $f_0$, directly influencing the output signal strength. Both parameters can be analytically derived, which are given by Eqs. (9) and (10), respectively.

$$f_0 = \frac{1}{2\pi}\sqrt{\frac{R_1 + R_2}{R_1 R_2 R_3 C_1 C_2}} \tag{9}$$

$$G = 20\log_{10}\left(\frac{C_2 R_3}{R_1(C_1 + C_2)}\right) \tag{10}$$

Based on Eqs. (9) and (10), the true causal graph of this system is illustrated in Fig. 21. It can be observed that $R_2$ has no causal effect on $G$, indicating that not all component degradation in the circuit impacts system performance. This characteristic makes this system a suitable benchmark for evaluating the ability of causal discovery methods to correctly identify causal relationships between degradation paths.

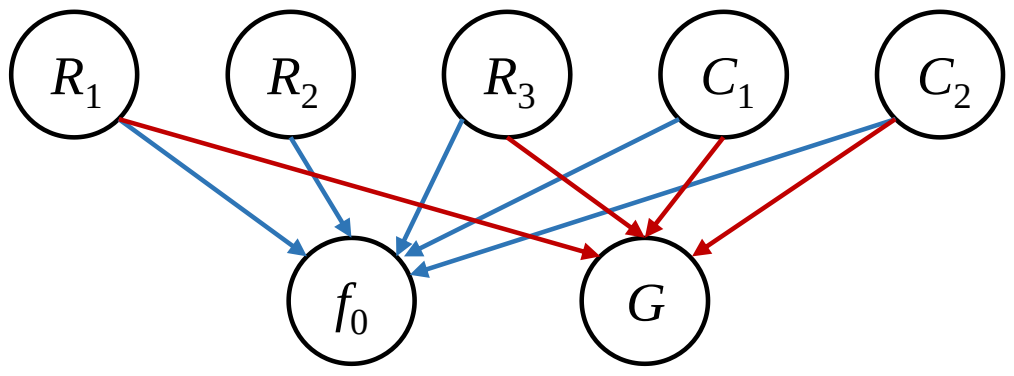

Fig. 21 True causal graph of the second-order multiple feedback band pass filter.

In real-world scenarios, resistance increases over time due to electro migration and material oxidation, while capacitance decreases due to electrode corrosion. Assuming that the operational amplifier is ideal, resistors degrade by an average of 20% and capacitors by 30% over a ten-year period, and the degradation processes of resistors and capacitors are linear [51]. Then, using the degradation model described by Eq. (4), the parameters for each component are listed in Table 8. According to Eqs. (9) and (10), the center frequency and peak gain of the initial circuit design are 16.78 kHz and 9.54 dB, respectively.

Table 8 Parameter settings for the second-order multiple feedback band pass filter.

| Parameters | $R_1$ (Ω) | $R_2$ (Ω) | $R_3$ (Ω) | $C_1$ (F) | $C_2$ (F) |
|---|---|---|---|---|---|
| $\mu_{X_0}$ | 5000 | 15000 | 25000 | 8e-10 | 1.2e-9 |
| $\sigma_{X_0}$ | 250 | 750 | 1250 | 8e-11 | 1.2e-10 |
| $\mu_a$ | 100 | 300 | 500 | -2.4e-11 | -3.6e-11 |
| $v_a$ | 0.01 | 0.01 | 0.01 | 0.01 | 0.01 |
| $\sigma$ | 100 | 300 | 500 | 1.6e-11 | 2.4e-11 |
| $\sigma_\varepsilon$ | 25 | 75 | 125 | 4e-12 | 6e-12 |
| $\gamma$ | 1 | 1 | 1 | 1 | 1 |

In this case, 20 degradation samples of the system were generated, with parameter values recorded every half a year. The values of the electrical components were obtained directly via Monte Carlo simulation based on their degradation models, and the corresponding center frequency and peak gain were obtained using a nonparametric frequency response estimation method. To achieve this, a simulation model was constructed in MATLAB/Simscape, as shown in Fig. 22. A linear chirp signal ranging from 10 kHz to 30 kHz was used as the input for frequency sweep. For each measurement time, the output of the filter was recorded, and the center frequency and peak gain were then determined by identifying the frequency at which the magnitude of the frequency response reached its maximum. The recorded degradation data are illustrated in Fig. 23.

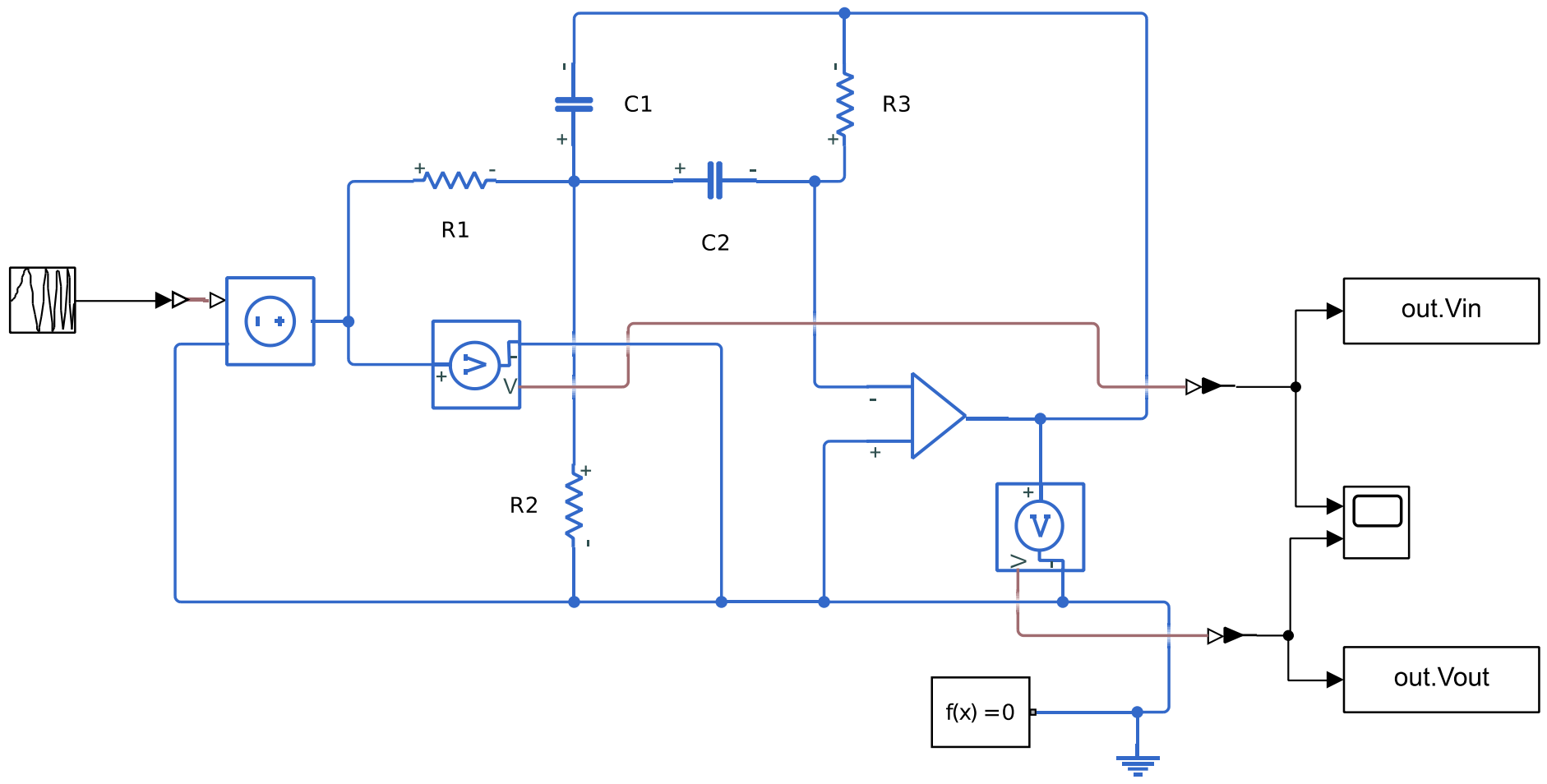

Fig. 22 Simulation model of the second-order multiple feedback band pass filter in MALAB/Simscape.

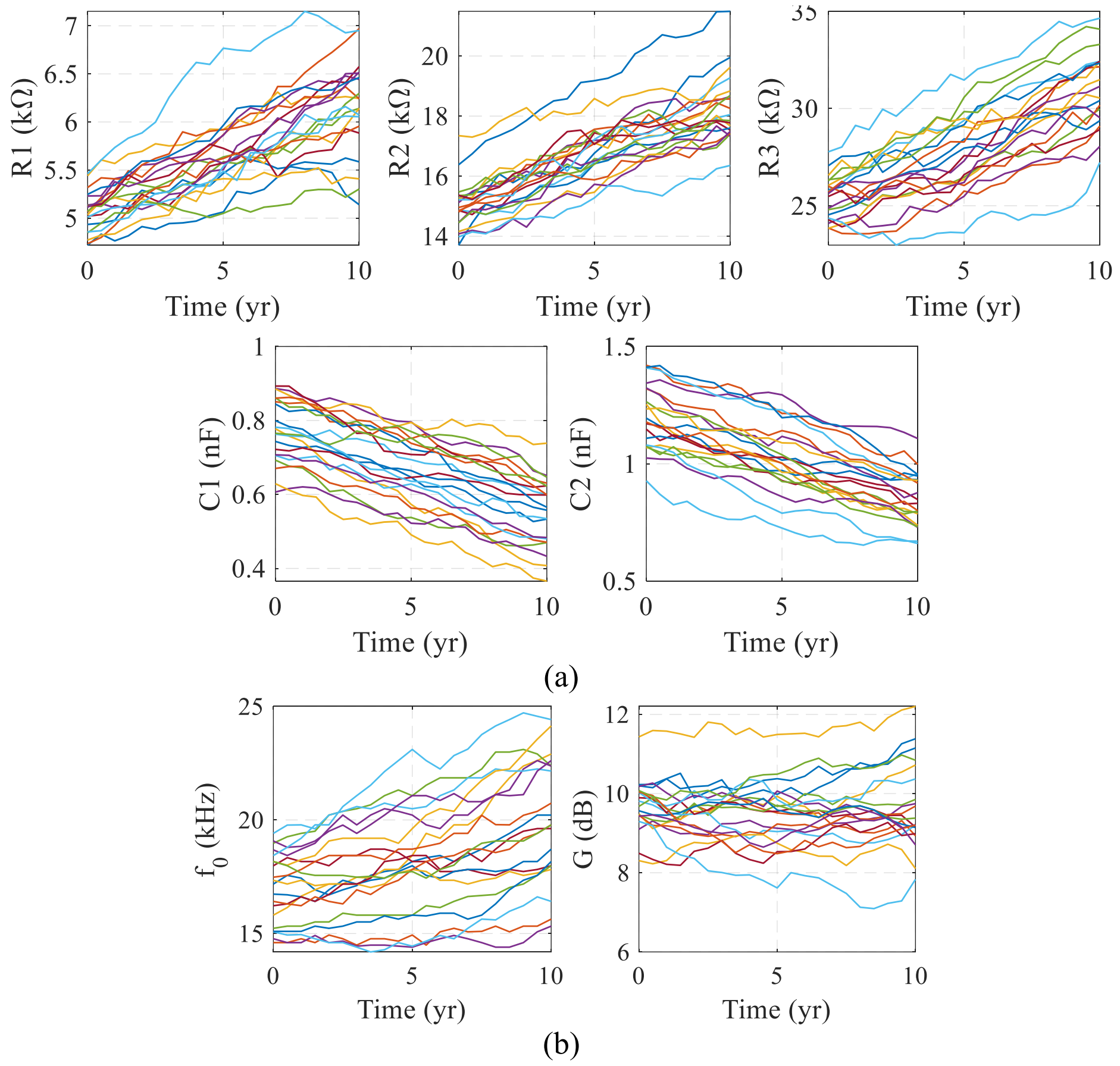

Fig. 23 Recorded degradation data of the second-order multiple feedback band pass filter: (a) Degradation of the circuit components; (b) Degradation of the system performance.

4.1.2 Results and analysis

Based on the degradation data given in Fig. 23, the results of different causal discovery techniques using strategy $S_2$ are given in Fig. 24. Notably, an undirected edge in the causal graph represents a causal link with an unknown direction. It can be deduced that:

- Stable-PC and GES correctly identify the causal skeleton but fail to determine the direction of

causal relationships;

- Direct-LiNGAM accurately identifies all independent degradation paths, but fails to fully recognize the paths with causal connections. Besides, the directions of causal relationships are always wrong.
- NOTEARS-Linear, NOTEARS-MLP and CaPS incorrectly infer causal relationships for independent degradation paths, and also fail to accurately identify the relationships of causally dependent degradation paths.
- Among all techniques, Stable-PC and GES deliver the most accurate causal discovery results. Although they lack the capacity to recognize causal directionality, domain knowledge can assist in clarifying and confirming the directionality of causal relationships.

Additionally, compared to the performance of causal discovery techniques on the numerical case in Table 7, Stable-PC, GES and Direct-LiNGAM exhibit similar performance on the filter case, whereas the other methods show substantial performance variation. Specifically, the accuracy of NOTEARS-Linear and NOTEARS-MLP declines considerably in identifying both independent and causally dependent degradation paths. CaPS also presents a substantial decrease in accuracy when inferring causal directions. This decline in performance may be attributed to the greater scale disparity among performance parameters in practical cases, as opposed to the more uniform magnitudes used in the numerical example.

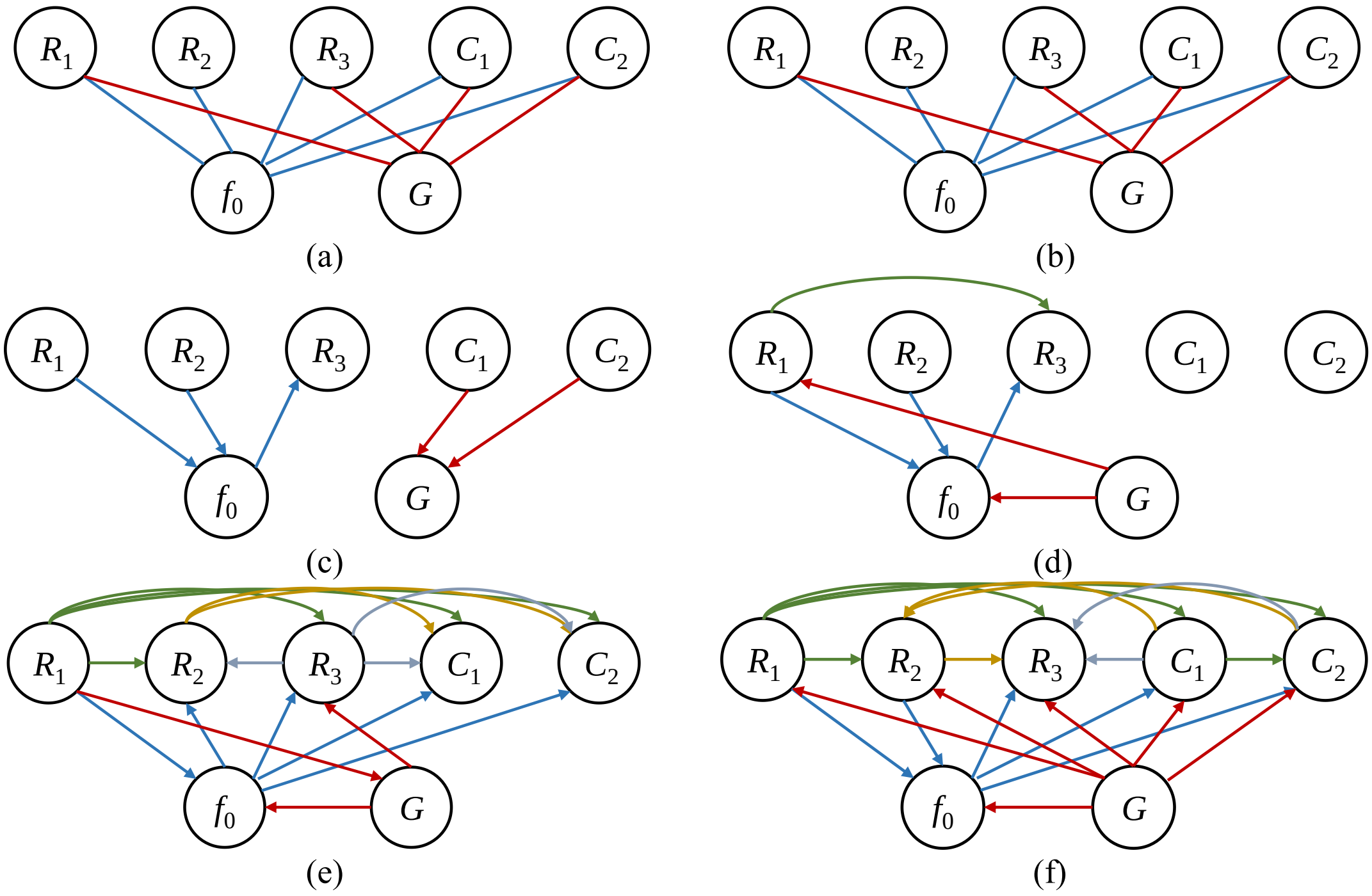

Fig. 24 Causal discovery results for the second-order multiple feedback band pass filter case using strategy $S_2$: (a) Stable-PC (b) GES (c) Direct-LiNGAM (d) NOTEARS-Linear (e) NOTEARS-MLP (f) CaPS.

Furthermore, to show the superiority of the proposed causal discovery strategy $S_2$, the results of different causal discovery techniques using strategy $S_1$ are given in Fig. 25. It is evident that under strategy $S_1$, all causal discovery techniques fail to identify the correct causal structure, especially in identifying independent degradation paths. This is due to the trend present in the degenerate data, which violates the IID assumption of non-temporal causal discovery techniques.

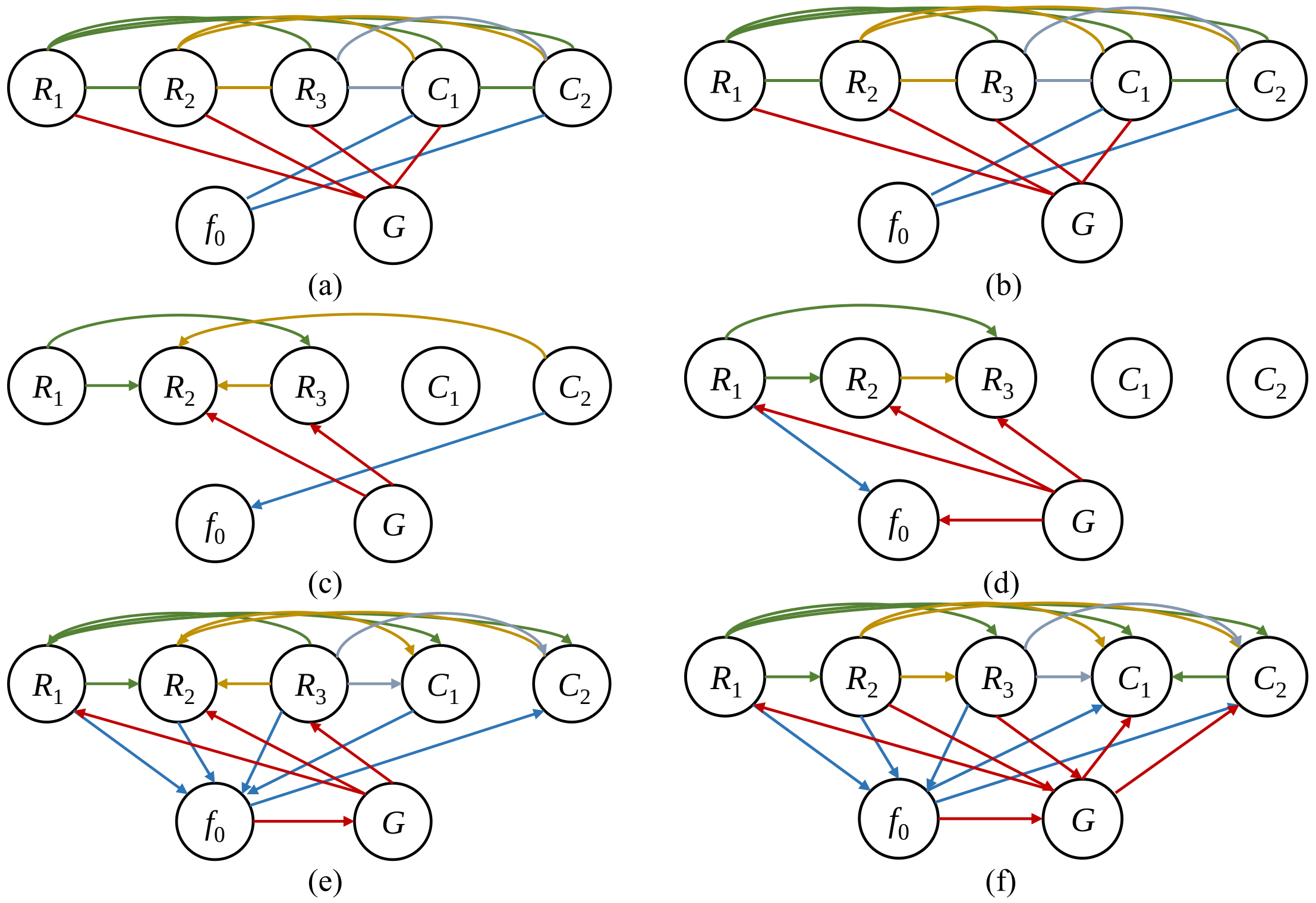

Fig. 25 Causal discovery results for the second-order multiple feedback band pass filter case using strategy $S_1$: (a) Stable-PC (b) GES (c) Direct-LiNGAM (d) NOTEARS-Linear (e) NOTEARS-MLP (f) CaPS.

## 4.2 A turbofan engine

### 4.2.1 Dataset description

In this section, the engineering feasibility of the proposed approach is illustrated through the C-MAPSS dataset [4]. This dataset is provided by the National Aeronautics and Space Administration (NASA)'s Ames Research Center, containing sensor measurements from multiple turbofan engines collected during their operational cycles.

Fig. 26 shows a diagram of a turbofan engine, consisting of a fan, low-pressure compressor (LPC), high-pressure compressor (HPC), combustor, high-pressure turbine (HPT), low-pressure turbine (LPT) and a nozzle. The dataset includes measurements of 21 sensors (i.e., performance parameters), of which the values for sensors 1, 5, 6, 10, 16, 18 and 19 are constant. Therefore, 14 performance parameters reflecting performance degradation over time are analyzed, as presented in Table 9.

The C-MAPSS dataset includes four distinct subsets: FD001, FD002, FD003 and FD004. Each dataset has measurements under different operational conditions and failure modes. In these datasets, engine units in the training set operate until they fail, while those in the test set are stopped prior to failure.

In this section, the training set of FD003 is employed for the case study, which includes sensor data from 100 different engine units. The engines in FD003 share identical operating conditions, characterized by altitude, Mach number and throttle resolver angle, and experience both fan and HPC degradation. All engines begin in a healthy state before performance degradation begins, with the degradation starting at random cycles. To ensure data extraction in the degrading phase, the last 40 cycles of sensor readings from each engine are selected for causal discovery, with an example of Ps30 shown in Fig. 27. Furthermore, to ensure the robustness of the causal discovery results, the analysis is repeated 20 times, each time randomly selecting 80% of the engine samples.

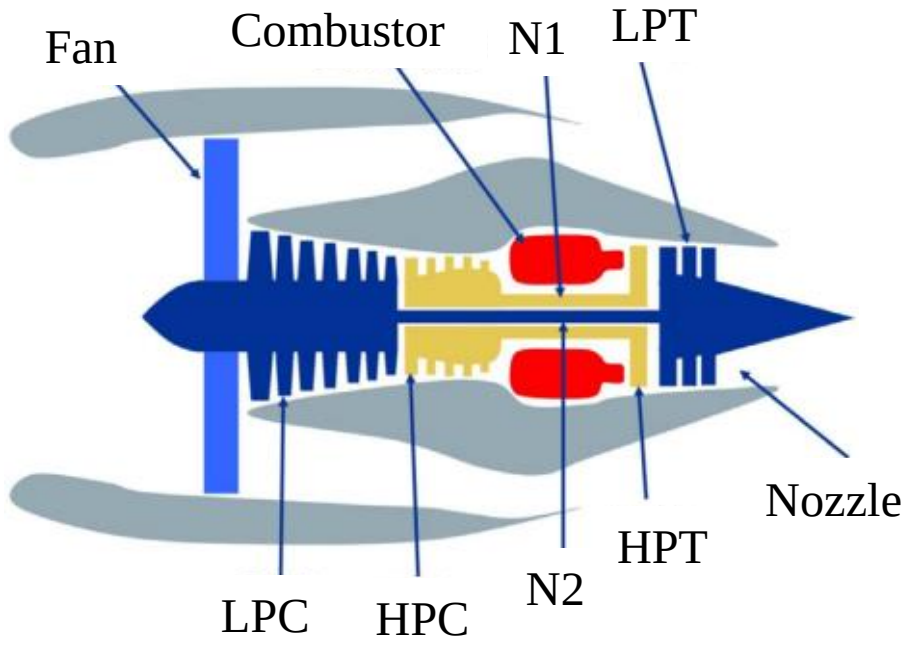

Fig. 26 Diagram of a turbofan engine [4].

Table 9 Performance recorded in the C-MAPSS dataset with degradation [4].

| Number | Symbols | Description | Unit |
|---|---|---|---|
| 1 | T24 | Total temperature at LPC outlet | °R |
| 2 | T30 | Total temperature at HPC outlet | °R |
| 3 | T50 | Total temperature at LPT outlet | °R |
| 4 | P30 | Total pressure at HPC outlet | psia |
| 5 | Nf | Physical fan speed | rpm |
| 6 | Nc | Physical core speed | rpm |
| 7 | Ps30 | Static pressure at HPC outlet | psia |
| 8 | phi | Ratio of fuel flow to Ps30 | pps/psi |
| 9 | NRf | Corrected fan speed | rpm |
| 10 | NRc | Corrected core speed | rpm |
| 11 | BPR | Bypass Ratio | / |
| 12 | htBleed | Bleed Enthalpy | / |
| 13 | W31 | HPT coolant bleed | lbm/s |
| 14 | W32 | LPT coolant bleed | lbm/s |

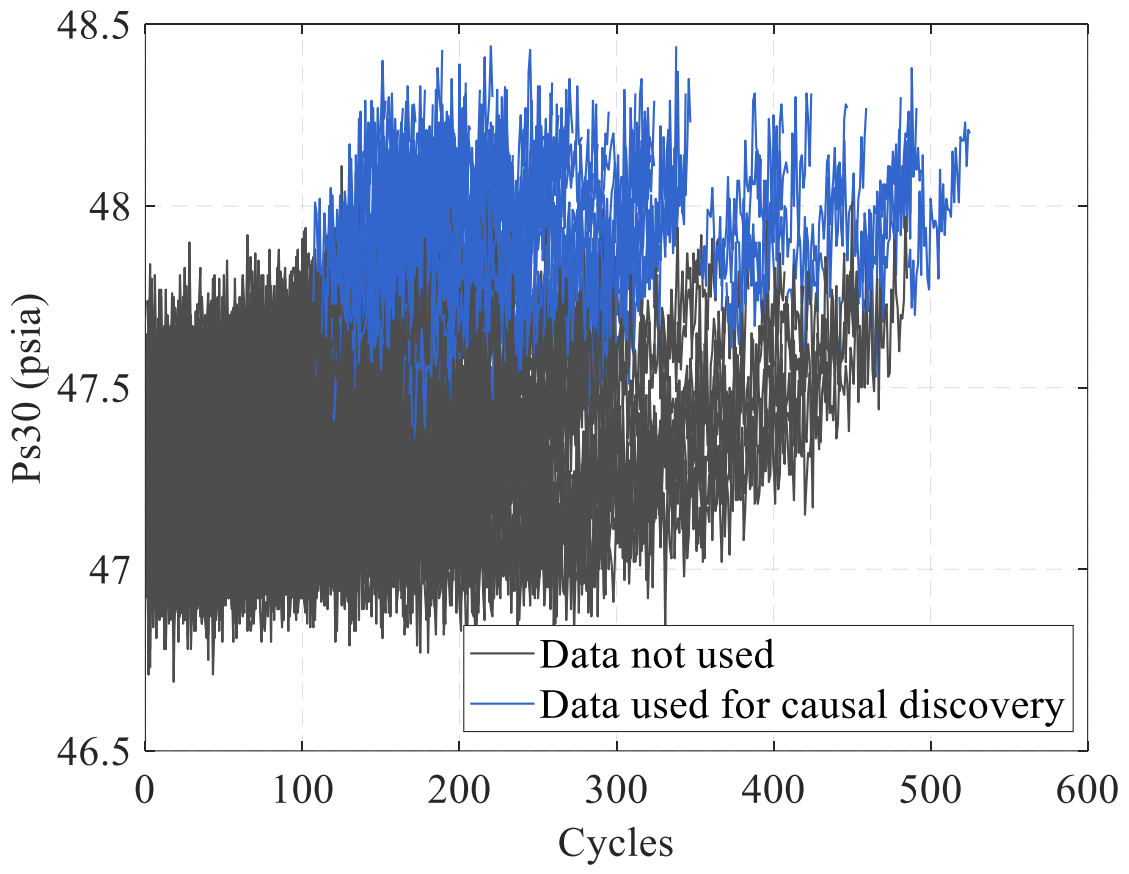

Fig. 27 Degradation data extraction of Ps30.

4.2.2 Results and analysis

Based on the C-MAPSS dataset, the results of different causal discovery techniques using strategy $S_2$ are given in Fig. 28. Similarly, an undirected edge in the causal graph represents a causal link with an unknown direction. It can be illustrated that Stable-PC and GES give robust causal skeleton identification but fail to determine the direction of causal relationships. In contrast, Direct-LiNGAM, NOTEARS-Linear, NOTEARS-MLP and CaPS fail to give effective causal discovery results.

Specifically, Stable-PC uncovers 18 causal links, and GES finds 12. Notably, each of the causal links identified by GES is included among those identified by Stable-PC. The effectiveness of each causal link is analyzed in Table 10. It can be seen that most of the identified causal links align with domain knowledge, demonstrating the effectiveness of the approach.

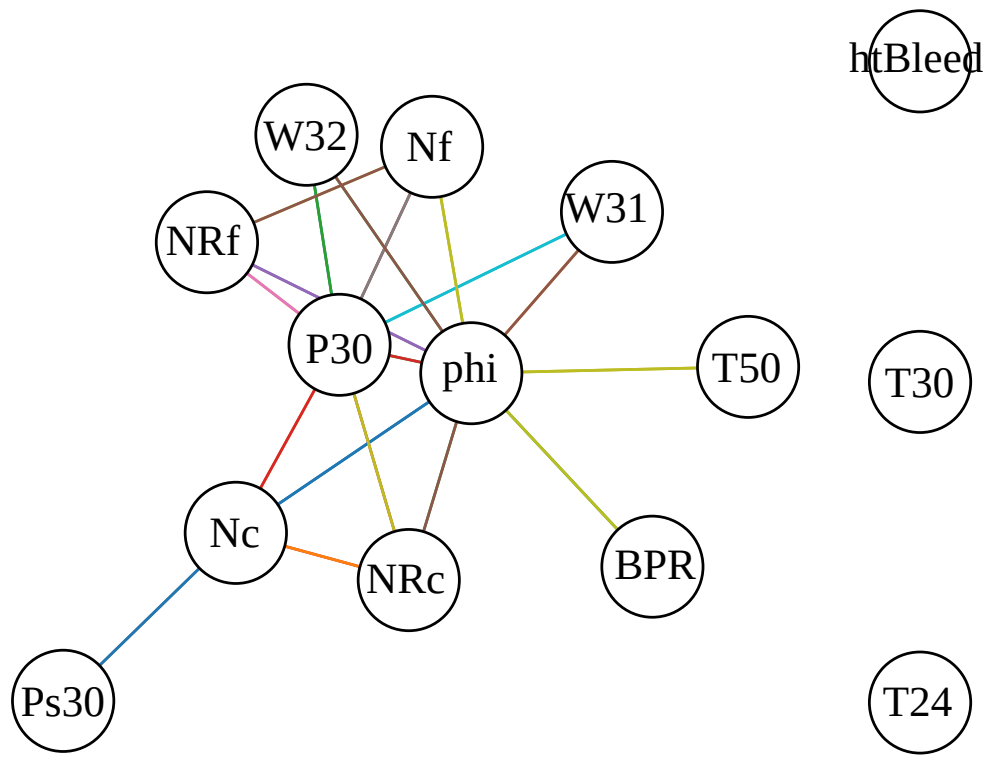

(a)

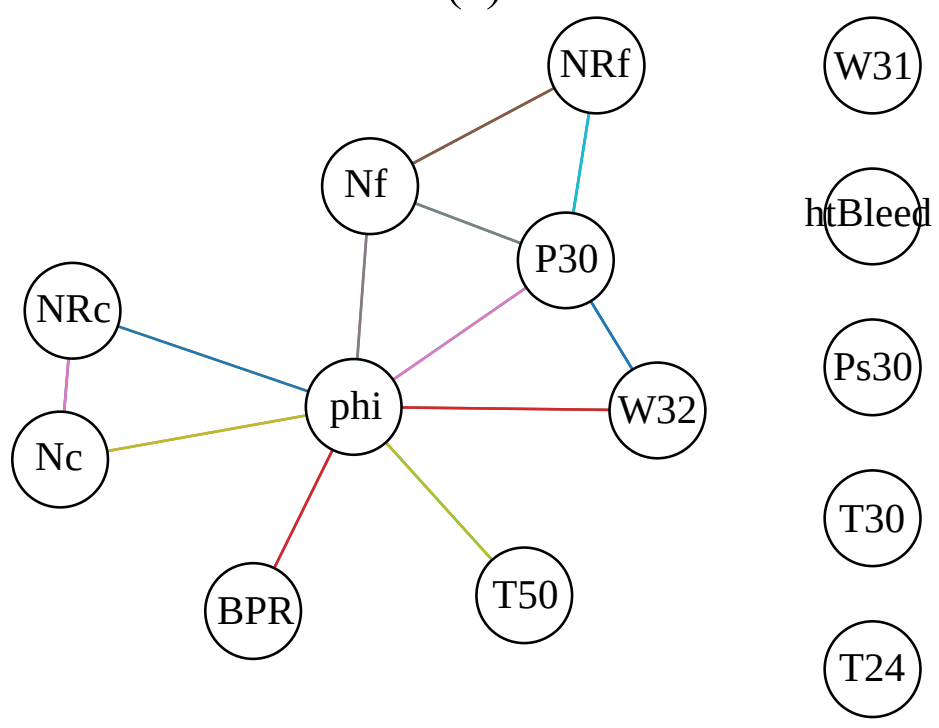

(b)

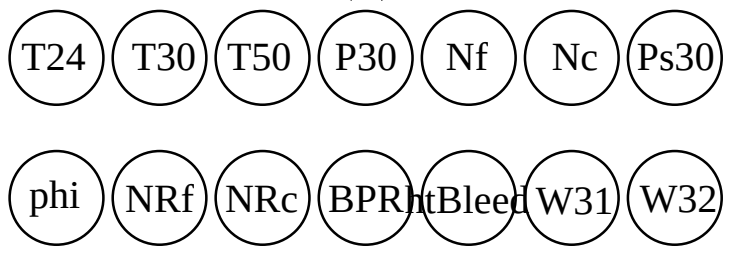

(c)

Fig. 28 Causal discovery results for the turbofan engine case using strategy $S_2$: (a) Stable-PC (b) GES (c) Direct-LiNGAM, NOTEARS-Linear, NOTEARS-MLP and CaPS.

Table 10 Effectiveness analysis of the causal links identified by Stable-PC and GES.

| Causal Link | Stable-PC | GES | Domain Knowledge | Explanation |
|---|---|---|---|---|
| Nf - NRf | √ | √ | E | NRf is the corrected value of Nf |
| Nc - NRc | √ | √ | E | NRc is the corrected value of Nc |
| P30 - phi | √ | √ | P | phi is defined as the ratio of fuel flow to Ps30, where Ps30 is closely related to P30 via flow velocity and fluid dynamic principles |
| phi - Nf | √ | √ | E | An increase in phi enhances combustion energy, which increases turbine work and subsequently raises Nf |
| phi - Nc | √ | √ | E | An increase in phi intensifies combustion, thereby altering the thermodynamic state of the core gas path. The enhanced energy flow through the HPT results in an increase in Nc |
| phi - NRf | √ | | E | NRf is the corrected value of Nf |
| phi - NRc | √ | √ | E | NRc is the corrected value of Nc |
| BPR - phi | √ | √ | P | A drop in BPR may lead to an increase in phi as a compensatory response, enabling a rise in Nc through boosted combustion energy |
| phi - W32 | √ | √ | P | phi influences T50 by increasing combustion energy, and W32 is adjusted in response to T50 to manage thermal load |
| phi - W31 | √ | | P | phi influences total temperature at HPT outlet by increasing combustion energy, and W31 is adjusted in response to T50 to manage thermal load |
| phi - T50 | √ | √ | E | Higher phi leads to stronger combustion, resulting in increased T50 |
| P30 - W32 | √ | √ | P | An increase in P30 provides a higher pressure head at the HPC outlet, improving the capability to divert air for LPT cooling, thereby potentially increasing W32 |
| P30 - W31 | √ | | P | An increase in P30 provides a higher pressure head at the HPC outlet, improving the capability to divert air for HPT cooling, thereby potentially increasing W31 |
| P30 - Nf | √ | √ | N | Changes in P30 reflect core compression dynamics, whereas Nf is determined by LPT power. Therefore, no direct causal link exists between them |
| P30 - NRf | √ | √ | N | NRf is the corrected value of Nf |
| Nc - P30 | √ | | E | An increase in Nc enhances the ability of the compressor to raise pressure, thereby increasing P30 |
| NRc - P30 | √ | | P | NRc is the corrected value of Nc |
| Nc - Ps30 | √ | | E | An increase in Nc represents faster core shaft rotation, improving the compression ratio of the HPC and raises Ps30 |

*Note*: Based on domain knowledge, “E” indicates that a causal link is known to exist, “P” represents that the causal link possibly exists, and “N” suggests that no causal link exists.

### 4.3 Merits of causal graphs between degradation paths

Causal graphs between degradation paths can provide substantial value for both degradation modeling and degradation control, particularly in complex engineering systems. The practical benefits of utilizing these causal graphs are outlined below:

**Enhanced degradation modeling**: Traditional degradation modeling typically focuses on directly modeling key performance indicators (KPIs) without considering the influence of other internal parameters. However, the degradation of KPIs often results from the degradation of multiple internal components, governed by inherent physical and functional dependencies. These dependencies introduce complexities in the degradation patterns of KPIs, making it difficult to describe the degradation trajectory accurately by a standard time-scale function. By employing causal graphs, the degradation of KPIs can be modeled more precisely by using the degradation paths of their causal ancestors as inputs.

**Informed degradation control**: Causal graphs between degradation paths clearly show how degradation of internal parameters directly impacts KPI degradation. This insight is valuable for prioritizing monitoring and control efforts toward the impactful components. For example, in the second-order multiple-feedback band pass filter case study, the causal graph illustrates that not all components in the circuit contribute to the degradation of peak gain. Recognizing such distinctions allows for more targeted maintenance strategies, helping to mitigate the impact of critical component degradation on the KPI of interest.

## 5 Conclusions and future directions

Capturing the dependencies between degradation paths is crucial for precise degradation modeling and effective degradation control. This study aims to explore effective strategies and techniques for identifying causal relationships between degradation paths. A causal discovery strategy based on degradation increments is proposed to mitigate the violation of the IID assumption when using raw degradation data. Through extensive numerical experiments and engineering cases, the effectiveness of six non-temporal causal discovery methods are compared. The findings of this study can be summarized as follows:

- Directly applying non-temporal causal discovery techniques to raw degradation data (strategy $S_1$) leads to incorrect identification of causal relationships due to the inherent trend in degradation processes. In contrast, the proposed strategy based on degradation increments (strategy $S_2$) effectively mitigates this issue and improves causal discovery accuracy.
- Among all tested techniques, Stable-PC and GES exhibit robust and accurate performance across both numerical and engineering cases, which are recommended for causal discovery between degradation paths. Notably, Stable-PC demonstrates superior performance in the turbine engine case by uncovering more effective causal links. Although these two methods lack the capacity to recognize causal directionality, domain knowledge can assist in clarifying and confirming the directionality of causal relationships.
- NOTEARS-MLP exhibits the highest performance in the numerical study, requiring nine samples to achieve accurate and stable results in most cases considered for both independent and causally dependent degradation paths. However, its performance declines considerably in the engineering cases, which may be attributed to the greater scale disparity among performance parameters in practical cases.
- Sensitivity analysis demonstrates that accurate identification of relationships for both independent and causally dependent degradation paths requires strong causal relationships, significant degradation trends, low measurement errors and large diffusion coefficients.

Based on the work in this study, several important directions can be extended. First, this study focuses on pairwise causal discovery. Future research should investigate causal discovery among multiple degradation paths to understand comprehensive causal relationships in complex systems. Besides, future work should investigate how causal discovery results can be effectively incorporated into multivariate degradation models to enhance reliability modeling, RUL prediction and maintenance decision-making in engineering applications.

## Declarations of interest

None.

## Acknowledgements

This work was supported by the China Scholarship Council [grant number 202406020189], the National Natural Science Foundation of China [grant number 51775020], the Science Challenge Project [grant number. TZ2018007], and the National Natural Science Foundation of China [grant numbers 62073009].